\documentclass[12pt,a4paper,aps,preprint,superscriptaddress,nofootinbib]{revtex4-1}
\usepackage[utf8]{inputenc}
\usepackage{graphicx}
\usepackage{amssymb}
\usepackage{textcomp}
\usepackage{amsmath}
\usepackage{tabularx}
\usepackage{bm}
\usepackage{times}
\usepackage{slashed}
\usepackage{multirow}
\usepackage{verbatim}
\usepackage{cancel}
\usepackage{subfigure}
\usepackage{ulem}

\usepackage[colorlinks=true, pdfstartview=FitV, linkcolor=blue, citecolor=blue, urlcolor=blue]{hyperref}
\allowdisplaybreaks[4]

\def\lsim{\mathrel{\raise.3ex\hbox{$<$\kern-.75em\lower1ex\hbox{$\sim$}}}}
\def\gsim{\mathrel{\raise.3ex\hbox{$>$\kern-.75em\lower1ex\hbox{$\sim$}}}}

\begin{document}

\title{Dark Photon Dark Matter in Quantum Electromagnetodynamics and Detection at Haloscope Experiments}

\author{Chang-Jie Dai}
\email{daichangjie@mail.nankai.edu.cn}
\affiliation{
School of Physics, Nankai University, Tianjin 300071, China
}

\author{Tong Li~\footnote[1]{corresponding author}}
\email{litong@nankai.edu.cn}
\affiliation{
School of Physics, Nankai University, Tianjin 300071, China
}

\author{Rui-Jia Zhang}
\email{zhangruijia@mail.nankai.edu.cn}
\affiliation{
School of Physics, Nankai University, Tianjin 300071, China
}

\begin{abstract}
The ultralight dark photon is one of intriguing dark matter candidates.
The interaction between the visible photon and dark photon is introduced by the gauge kinetic mixing between the field strength tensors of the Abelian gauge groups in the Standard Model and dark sector. The relativistic electrodynamics was generalized to quantum electromagnetodynamics (QEMD) in the presence of both electric and magnetic charges. The photon is described by two four-potentials corresponding to two $U(1)$ gauge groups and satisfying non-trivial commutation relations.
In this work, we construct the low-energy dark photon-photon interactions in the framework of QEMD and obtain new dark photon-photon kinetic mixings. The consequent field equations and the new Maxwell's equations are derived in this framework. We also investigate the detection strategies of dark photon as light dark matter as well as the generic kinetic mixings at haloscope experiments.
\end{abstract}

\maketitle

\section{Introduction}
\label{sec:Intro}

The numerous candidates of dark matter (DM) motivate us to search for potential hidden particles in a wide range of mass scale.
The dark photon (DP, or called hidden photon)~\cite{Holdom:1985ag,Holdom:1986eq} is an appealing candidate of ultralight bosonic DM~\cite{Nelson:2011sf,Arias:2012az,Graham:2015rva} (see a recent review Ref.~\cite{Fabbrichesi:2020wbt} and references therein). It is a spin-one field particle gauged by an Abelian group in dark sector. The interaction between the visible photon and the dark photon is through the gauge kinetic
mixing between the field strength tensors of the Standard Model (SM) electromagnetic gauge group $U(1)_{\rm EM}$ and the dark Abelian gauge group $U(1)_{\rm D}$ below the electroweak scale
\begin{eqnarray}
\mathcal{L}\supset -\frac{1}{4}F^{\mu\nu}F_{\mu\nu}-\frac{1}{4}{F}_D^{\mu\nu}{F}_{D\mu\nu}-\frac{\epsilon}{2}F^{\mu\nu}{F}_{D\mu\nu}+\frac{1}{2}m_{D}^2{A_D}^\mu A_{D\mu}\;,
\end{eqnarray}
where $F^{\mu\nu}$ ($F^{\mu\nu}_D$) is the SM (dark) field strength, and $A_D$ is the dark gauge boson with mass $m_{D}$. Suppose the SM particles are uncharged under the dark gauge group, this kinetic mixing $\epsilon\ll 1$ is generated by integrating out new heavy particles charged under both gauge groups at loop level. The two gauge fields can be rotated to get rid of the mixing and as a result, the SM matter current gains a shift by $A_\mu\to A_\mu - \epsilon A_{D\mu}$.
Based on the framework of quantum electrodynamics (QED), the electromagnetic signals from the source of dark photon DM can be searched for in terrestrial experiments~\cite{Arias:2012az,Arias:2014ela,Chaudhuri:2014dla,Nguyen:2019xuh,Cervantes:2022epl,Cervantes:2022gtv,McAllister:2022ibe,Cervantes:2022yzp,Ramanathan:2022egk,Schneemann:2023bqc,Tang:2023oid,He:2024ytp}.

The description of relativistic electrodynamics may not be as simple as QED theory. The magnetic monopole is one of the most longstanding and mysterious topics in history~\cite{Dirac:1931kp,Wu:1975es,tHooft:1974kcl,Polyakov:1974ek,Cho:1996qd,Hung:2020vuo,Alexandre:2019iub,Ellis:2016glu,Lazarides:2021los}.
In 1960's, J. S. Schwinger and D. Zwanziger developed a generalized electrodynamics with monopoles in the presence of both electric and magnetic charges, called
quantum electromagnetodynamics (QEMD)~\cite{Schwinger:1966nj,Zwanziger:1968rs,Zwanziger:1970hk}. The characteristic feature of QEMD is to substitute the $U(1)_{\rm EM}$ gauge group by two $U(1)$ gauge groups to introduce both electric and magnetic charges. Two four-potentials $A_\mu$ and $B_\mu$ (instead of only one $A_\mu$ in QED) are introduced corresponding to
the two $U(1)$ gauge groups (called $U(1)_{A}$ and $U(1)_B$ below), respectively.
They formally built a local Lagrangian density, a non-trivial form of equal-time canonical commutation relations and resulting Lagrangian field equations in a local quantum field theory. Zwanziger et al. also proved that the right degrees of freedom of physical photon are preserved and the Lorentz invariance is not
violated in this theory~\cite{Brandt:1977be,Brandt:1978wc,Sokolov:2023pos}. Recently, based on the framework of QEMD, it was pointed out that more generic
axion-photon interactions may arise and there appeared quite a few studies of them in theory~\cite{Sokolov:2022fvs,Sokolov:2023pos,Heidenreich:2023pbi,Li:2023zcp} and phenomenology~\cite{Li:2022oel,Tobar:2022rko,McAllister:2022ibe,Li:2023kfh,Li:2023aow,Tobar:2023rga,Patkos:2023lof,Dai:2024dkr}. As a result of axion-monopole dynamics, more anomalous axion-photon interactions and couplings arise in contrary to the ordinary axion-photon coupling. Our previous work investigated the new axion-modified Maxwell equations and analytically obtained the axion-induced electromagnetic fields~\cite{Li:2022oel}. Based on the solutions, we proposed new strategies to probe the new couplings of axion in LC circuit experiment~\cite{Li:2022oel}, cavity experiment~\cite{Li:2023aow}, interface haloscope experiment~\cite{Li:2023kfh} and superconducting radio frequency cavity experiment~\cite{Dai:2024dkr}. This article aims to extend the closely related study to the field of DP in QEMD.

In this work, we construct the dark photon-photon interactions in the framework of QEMD and investigate the relevant detection strategies of light dark photon DM. We introduce new heavy fermions $\psi$ charged under the four electromagnetic $U(1)$ groups, i.e., $U(1)_{A}\times U(1)_B$ in visible sector and $U(1)_{A_D}\times U(1)_{B_D}$ in dark sector. The covariant
derivative of $\psi$ fermion in the kinetic term is then given by
\begin{eqnarray}
i\Bar{\psi}\gamma^\mu D_\mu \psi = i\Bar{\psi}\gamma^\mu (\partial_\mu- e q_\psi A_\mu - g g_\psi B_\mu - e q_{D\psi} A_{D\mu} - g g_{D\psi} B_{D\mu})\psi\;,
\end{eqnarray}
where $A_\mu, B_\mu$ ($A_{D\mu}, B_{D\mu}$) are the potentials in the visible (dark) sector, and they are multiplied by the corresponding electric and magnetic charges.
The visible photon is described by the two four-potentials $A_\mu, B_\mu$ in the visible sector, and the dark photon is gauged under either QED (with only $A_{D\mu}$) or QEMD (with both $A_{D\mu}$ and $B_{D\mu}$) in dark sector.
After integrating out the new fermions $\psi$ in vacuum polarization diagrams of the four potentials, one can obtain the new kinetic mixings between dark photon and visible photon. We show their low-energy Lagrangian and the consequent field equations which are equivalent to new Maxwell's equations of dark photon. Based on the new Maxwell's equations, we also study the detection strategies through haloscope experiments to search for the light dark photon DM in this framework as well as the new kinetic mixings.

There were literatures also studying the kinetic mixing between two Abelian gauge theories that have both electric and magnetic charges~\cite{Hook:2017vyc,Terning:2018lsv,Terning:2019bhg,Terning:2020dzg}. However, besides the ordinary kinetic mixing of the visible photon to the dark photon $F_{\mu\nu}F_D^{\mu\nu}$ that they focused, in this work we build a complete low-energy Lagrangian of two Abelian gauge fields in both visible and dark sectors and introduce an additional kinetic mixing between dark photon and visible photon. The complete low-energy Lagrangian with this new kinetic mixing and a two-component DP DM scenario induce intriguing phenomenologies as we will discuss below.

This paper is organized as follows. In Sec.~\ref{sec:QEMD}, we introduce the QEMD theory and the effective Lagrangian of dark photon and visible photon in QEMD framework. In Sec.~\ref{sec:Maxwell}, we show the generic kinetic mixing terms in the Lagrangian. The consequent field equations and the new Maxwell's equations are then derived in this framework. We discuss the setup and signal power of haloscope experiments for the generic kinetic mixings and dark photon DM in Sec.~\ref{sec:Sens}. We also show the sensitivity of haloscope experiments to each kinetic mixing or dark photon DM component. Our conclusions are drawn in Sec.~\ref{sec:Con}.

\section{Formalism of photon and dark photon in QEMD framework}
\label{sec:QEMD}

In this section, we first describe the framework of QEMD theory and then introduce the necessary ingredients for constructing the extended dark photon-photon interactions based on QEMD.

To properly build a relativistic electrodynamics in the presence of magnetic monopole, a reliable method is to introduce two four-potentials $A_\mu$ and $B_\mu$ corresponding to two $U(1)$ gauge groups $U(1)_A$ and $U(1)_B$, respectively~\cite{Schwinger:1966nj,Zwanziger:1968rs,Zwanziger:1970hk}. Both the electric and magnetic charges are inherently brought into the same theoretical framework. The general Maxwell's equations in the presence of electric and magnetic currents are
\begin{eqnarray}
\partial_\mu F^{\mu\nu} = j_e^\nu\;,~~\partial_\mu F^{d~\mu\nu} = j_m^\nu\;,
\label{eq:FFd}
\end{eqnarray}
where the Hodge dual of field strength $F^{\mu\nu}$ is $F^{d~\mu\nu}={1\over 2}\epsilon^{\mu\nu\rho\sigma}F_{\rho\sigma}$ with $\epsilon_{0123}=+1$, and the currents are conserved with $\partial_\mu j_e^\mu=\partial_\mu j_m^\mu=0$. The general solutions to the above equations are
\begin{eqnarray}
F&=&\partial\wedge A-(n\cdot\partial)^{-1}(n\wedge j_m)^d\;,\\
F^d&=&\partial\wedge B+(n\cdot\partial)^{-1}(n\wedge j_e)^d\;,
\end{eqnarray}
where $n^\mu = (0,\vec{n})$ is an arbitrary space-like vector, the integral operator $(n\cdot\partial)^{-1}$ satisfies $n\cdot \partial (n\cdot\partial)^{-1}(x)=\delta^4(x)$ and we define $(X \wedge Y)^{\mu \nu}\equiv X^\mu Y^\nu - X^\nu Y^\mu$ for any four-vectors $X$ and $Y$.
The above field strength tensors satisfy
\begin{eqnarray}
n\cdot F = n\cdot (\partial\wedge A)\;,~~n\cdot F^d = n\cdot (\partial\wedge B)\;.
\end{eqnarray}
Using the identity $G=(1/n^2) [(n\wedge (n\cdot G))- (n\wedge (n\cdot G^d))^d]$ for any antisymmetric tensor $G$, one can rewrite $F$ and $F^d$ only in terms of potentials
\begin{eqnarray}
F&=&\frac{1}{n^2}(n\wedge[n\cdot(\partial\wedge A)]-n\wedge[n\cdot(\partial\wedge B)]^d)\;,\\
F^d&=&\frac{1}{n^2}(n\wedge[n\cdot(\partial\wedge A)]^d+n\wedge[n\cdot(\partial\wedge B)])\;.
\end{eqnarray}
After substituting them into Eq.~(\ref{eq:FFd}), we obtain the Maxwell's equations
\begin{eqnarray}
\frac{n\cdot\partial}{n^2}(n\cdot\partial A^\mu-\partial^\mu n\cdot A-n^\mu \partial\cdot A-\epsilon^\mu_{\nu\rho\sigma}n^\nu\partial^\rho B^\sigma)&=&j_e^\mu\;,\\
\frac{n\cdot\partial}{n^2}(n\cdot\partial B^\mu-\partial^\mu n\cdot B-n^\mu\partial\cdot B+\epsilon^\mu_{\nu\rho\sigma}n^\nu\partial^\rho A^\sigma)&=&j_m^\mu\;.
\end{eqnarray}

These Maxwell's equations can be realized by the local Lagrangian of photon as follows~\cite{Zwanziger:1970hk}
\begin{eqnarray}
\mathcal{L}_{\rm P}&=&-\frac{1}{2n^2}[n\cdot(\partial\wedge A)]\cdot[n\cdot(\partial \wedge B)^d]+\frac{1}{2n^2}[n\cdot(\partial\wedge B)]\cdot[n\cdot(\partial\wedge A)^d]\nonumber\\
&&-\frac{1}{2n^2}[n\cdot(\partial\wedge A)]^2-\frac{1}{2n^2}[n\cdot(\partial\wedge B)]^2-j_e\cdot A-j_m\cdot B+\mathcal{L}_G\;,
\label{eq:QEMDL}
\end{eqnarray}
where $\mathcal{L}_G=(1/2n^2)\{[\partial(n\cdot A)]^2+[\partial(n\cdot B)]^2\}$ is a gauge fixing term. One can rewrite it in terms of canonical variables and get the non-trivial commutation relations between the two four-potentials~\cite{Zwanziger:1970hk}
\begin{eqnarray}
\label{eq:commutation1}[A^\mu(t,\vec{x}),B^\nu(t,\vec{y})]&=&i\epsilon^{\mu\nu}_{~~\kappa 0} n^\kappa (n\cdot \partial)^{-1}(\vec{x}-\vec{y})\;,\\
~
\label{eq:commutation2}[A^\mu(t,\vec{x}),A^\nu(t,\vec{y})]&=&[B^\mu(t,\vec{x}),B^\nu(t,\vec{y})]=-i(g_0^{~\mu} n^\nu+g_0^{~\nu} n^\mu)(n\cdot \partial)^{-1}(\vec{x}-\vec{y})\;.
\end{eqnarray}
The right number of photon degrees of freedom is preserved due to the constraints from the above equations of motion, gauge condition and equal-time commutation relations.
We notice the other important identity between two antisymmetric tensors $G$ and $H$
\begin{eqnarray}
{\rm tr}(G\cdot H)=G^{\mu\nu}H_{\nu\mu} = {2\over n^2}[-(n\cdot G)(n\cdot H) + (n\cdot G^d)(n\cdot H^d)]\;.
\end{eqnarray}
The QEMD Lagrangian of visible photon is then rewritten as
\begin{eqnarray}
\mathcal{L}_{\rm P}&=&
\frac{1}{4}{\rm tr}(F\cdot (\partial\wedge A))
+\frac{1}{4}{\rm tr}(F^{d}\cdot (\partial\wedge B))-j_e\cdot A-j_m\cdot B+\mathcal{L}_G\;.
\label{eq:QEMDLre}
\end{eqnarray}
The same result can also be obtained based on Schwinger's phenomenological source theory (PST)~\cite{Schwinger:1966zz,Schwinger:1967rg}.
PST introduces source function to express the particles involved in a collision. The vacuum amplitude between two types of sources yields the $S$ matrix element. For the theory of magnetic charge, based on PST, Ref.~\cite{Schwinger:1968rq} showed the same action of photon as Eq.~(\ref{eq:QEMDLre}) in QEMD theory. Similarly, the Lagrangian of massive dark photon gains the following form
\begin{eqnarray}
\mathcal{L}_{\rm DP}&=&
-\frac{1}{2n^2}[n\cdot(\partial\wedge A_D)]\cdot[n\cdot(\partial \wedge B_D)^d]+\frac{1}{2n^2}[n\cdot(\partial\wedge B_D)]\cdot[n\cdot(\partial\wedge A_D)^d]\nonumber\\
&&-\frac{1}{2n^2}[n\cdot(\partial\wedge A_D)]^2-\frac{1}{2n^2}[n\cdot(\partial\wedge B_D)]^2+{1\over 2}m_D^2 A_D^\mu A_{D\mu}+{1\over 2}m_D^2 B_D^\mu B_{D\mu}+\mathcal{L}_{GD}\nonumber\\
&=&\frac{1}{4}{\rm tr}(F_D\cdot (\partial \wedge A_D))
+\frac{1}{4}{\rm tr}(F^{d}_D\cdot (\partial\wedge B_D))+{1\over 2}m_D^2 A_D^\mu A_{D\mu}+{1\over 2}m_D^2 B_D^\mu B_{D\mu}+\mathcal{L}_{GD}\;,
\end{eqnarray}
where $\mathcal{L}_{GD}$ denotes the gauge fixing term for DP.

The spatial vector $n_\mu$ introduced in the QEMD theory seems to violate the Lorentz invariance. This
originates from the non-locality of the QEMD theory.
Brandt, Neri and Zwanziger formally showed that the observables of the QEMD are Lorentz invariant using the path-integral approach~\cite{Brandt:1977be,Brandt:1978wc} (see also a recent demonstration in Ref.~\cite{Sokolov:2023pos}). They claimed that, after all the quantum
corrections are properly accounted for, the dependence on the spatial vector $n_\mu$ in the action factorizes into an integer linking number multiplied by the
combination of charges in the quantization condition $q_i g_j - q_j g_i$. This $n$ dependent part is then given by $2\pi$ multiplied by an integer. Since the action contributes to the generating functional in the exponential form, this
Lorentz-violating part does not play any role in physical processes. According to Refs.~\cite{PhysRev.150.1349,Schwinger:1968rq,Brandt:1978wc}, the kinetic term and the current terms can be rewritten as
\begin{eqnarray}
\mathcal{L}_{\rm P}\supset -\frac{1}{2}F\cdot(\partial\wedge A)+\frac{1}{4}F^2-j_e\cdot A -j_m\cdot B_n\;,
\end{eqnarray}
where the redefined potential is $B_n(x)=\int d\omega\cdot F^d(x-\omega)=(n\cdot\partial)^{-1}n\cdot F^d(x)$.
The action of QEMD theory remains invariant under the combined gauge transformation and Lorentz transformation~\cite{Brandt:1977be}
\begin{eqnarray}
F\rightarrow F,~~A_\mu\rightarrow A_\mu+\partial_\mu\lambda,~~B_n\rightarrow B_{n'}=(n'\cdot\partial)^{-1}n'\cdot F^d\;,
\label{eq:transformation}
\end{eqnarray}
where the function $\lambda(x)$ is determined by the condition
\begin{eqnarray}
    \partial\wedge\partial\lambda=\{[(n'\cdot\partial)^{-1}n'-(n\cdot\partial)^{-1}n]\wedge j_m\}^d\;.
\end{eqnarray}

The Lagrangian of massive DP gains two mass terms besides the conventional QEMD Lagrangian with the substitution of $A\to A_D$ and $B\to B_D$. The DP Lagrangian can be obtained by combining two forms of Lagrangian
\begin{eqnarray}
\mathcal{L}_{\rm DP1}&=&-\frac{1}{2}F_D\cdot(\partial\wedge A_D)+\frac{1}{4}F_D^2-j_{eD}\cdot A_D-j_{mD}\cdot B_{Dn}+m_D^2 A_D^2\;,\\
\mathcal{L}_{\rm DP2}&=&-\frac{1}{2}F_D^d\cdot(\partial\wedge B_D)+\frac{1}{4}F_D^{d~2}-j_{eD}\cdot A_{Dn}-j_{mD}\cdot B_D+m_D^2 B_D^2\;,
\end{eqnarray}
where $\mathcal{L}_{\rm DP1}$ ($\mathcal{L}_{\rm DP2}$) is composed of $F_D$, $A_D$ and $B_{Dn}$ ($F_D^d$, $B_D$ and $A_{Dn}$). Analogous to $\mathcal{L}_{\rm DP1}$, $\mathcal{L}_{\rm DP2}$ can also be proved to be Lorentz invariant~\cite{Schwinger:1968rq}. Thus, the DP Lagrangian with mass terms satisfy Lorentz symmetry. We omit the dark currents in the following calculation.

\section{Dark photon-photon interactions and field equations}
\label{sec:Maxwell}

Inspired by the two potential terms in either $\mathcal{L}_{\rm P}$ or $\mathcal{L}_{\rm DP}$, we can build the low-energy dark photon-photon kinetic mixing interactions as follows
\begin{align}
\mathcal{L}_{\rm DP-P}
=&\frac{\epsilon_{1}}{2}{\rm tr}(F\cdot(\partial\wedge A_D))
+\frac{\epsilon_{1}}{2}{\rm tr}(F^{d}\cdot (\partial\wedge B_D)) \notag\\
+&\frac{\epsilon_{2}}{2}{\rm tr}(F^{d}\cdot (\partial\wedge A_D))
-\frac{\epsilon_{2}}{2}{\rm tr}(F\cdot (\partial\wedge B_D)) \;.\label{eq:DP-P}
\end{align}
This Lagrangian is equivalent to the one with the substitution of $A \leftrightarrow A_D$ and $B\leftrightarrow B_D$. They contribute to the same equations of motion.
The two mixing parameters $\epsilon_1$ and $\epsilon_2$ can be obtained by integrating out the new heavy fermion $\psi$ in vacuum polarization diagram of the potentials in visible and dark sectors. Suppose the fermion $\psi$ is only charged in $U(1)_{A_D}$ group in dark sector, the above Lagrangian is simplified as
\begin{eqnarray}
\mathcal{L}^{~\prime}_{\rm DP-P}
=\frac{\epsilon_{1}}{2}{\rm tr}(F\cdot(\partial\wedge A_D))
+\frac{\epsilon_{2}}{2}{\rm tr}(F^{d}\cdot (\partial\wedge A_D))\;,
\end{eqnarray}
where the terms with potential $B_D$ in $\mathcal{L}_{\rm DP-P}$ vanish.

We next apply the Euler-Lagrange equation to the above DP-P Lagrangian $\mathcal{L}_{\rm DP-P}$ and then obtain the field equations of photon as
\begin{eqnarray}
&&\partial_{\mu}F^{\mu\nu}+\epsilon_{1}\partial_{\mu}F^{\mu\nu}_D-\epsilon_{2}\partial_{\mu}F^{d~\mu\nu}_D=j_e^{\nu} \;,\label{eq:Peq1}\\
&&\partial_{\mu}F^{d~\mu\nu}+\epsilon_{1}\partial_{\mu}F^{d~\mu\nu}_D+\epsilon_{2}\partial_{\mu}F^{\mu\nu}_D=j_m^{\nu}\;.
\label{eq:Peq2}
\end{eqnarray}
The equations of motion for dark photon are
\begin{eqnarray}
&&\partial_{\mu}F^{\mu\nu}_D+m_D^2A_D^{\nu}+\epsilon_{1}\partial_{\mu}F^{\mu\nu}+\epsilon_{2}\partial_{\mu}F^{d~\mu\nu}=0\;,
\\
&&
\partial_{\mu}F^{d~\mu\nu}_D+m_D^2B_D^{\nu}+\epsilon_{1}\partial_{\mu}F^{d~\mu\nu}-\epsilon_{2}\partial_{\mu}F^{\mu\nu}=0\;.
\end{eqnarray}
After inserting the dark photon equations into Eqs.~(\ref{eq:Peq1}) and (\ref{eq:Peq2}), we obtain the modified Maxwell's equations
\begin{eqnarray}
\partial_{\mu}F^{\mu\nu}=&\epsilon_{1}m_D^2A_D^{\nu}-\epsilon_{2}m_D^2B_D^{\nu}\;,
\label{eq:Maxeq1}\\ \partial_{\mu}F^{d~\mu\nu}=&\epsilon_{1}m_D^2B_D^{\nu}+\epsilon_{2}m_D^2A_D^{\nu}\;,
\label{eq:Maxeq2}
\end{eqnarray}
where the $\mathcal{O}(\epsilon^2_{1,2})$ terms are neglected, and the primary electromagnetic fields driven by the static currents and charges have been subtracted here.
Note that the right-handed sides of Eqs.~(\ref{eq:Maxeq1}) and (\ref{eq:Maxeq2}) are the linear combinations of $A_D$ and $B_D$. We perform an $O(2)$ transformation
\begin{eqnarray}
\left(
  \begin{array}{c}
    \tilde{A}_D \\
    \tilde{B}_D \\
  \end{array}
\right) =
\left(
  \begin{array}{cc}
    \cos\varphi & -\sin\varphi \\
    \sin\varphi & \cos\varphi \\
  \end{array}
\right)
\left(
  \begin{array}{c}
    A_D \\
    B_D \\
  \end{array}
\right)\;,
\end{eqnarray}
where $\cos\varphi=\epsilon_1/\sqrt{\epsilon_1^2+\epsilon_2^2}$ and $\sin\varphi=\epsilon_2/\sqrt{\epsilon_1^2+\epsilon_2^2}$. The equations then become
\begin{eqnarray}
\partial_{\mu}F^{\mu\nu}=&\epsilon m_D^2 \tilde{A}_D^{\nu}\;,\\
\partial_{\mu}F^{d~\mu\nu}=&\epsilon m_D^2 \tilde{B}_D^{\nu}\;,
\end{eqnarray}
where the only mixing parameter is $\epsilon=\sqrt{\epsilon_1^2+\epsilon_2^2}$.
Corresponding to the Lagrangian $\mathcal{L}_{\rm DP-P}^{~\prime}$, the above Maxwell's equations are simplified as
\begin{eqnarray}
\partial_{\mu}F^{\mu\nu}=&\epsilon_{1}m_D^2A_D^{\nu}\;,\\
\partial_{\mu}F^{d~\mu\nu}=&\epsilon_{2}m_D^2A_D^{\nu}\;,
\end{eqnarray}
where there is only $A_D$ in dark sector and the two equations rely on $\epsilon_1$ and $\epsilon_2$, respectively.

The modified Amp\`{e}re's law and Faraday's law equations become
\begin{eqnarray}
\vec{\nabla}\times\vec{\mathbb{B}}=\frac{\partial\vec{\mathbb{E}}}{\partial t}+\vec{j}_{eD}\;,\\
-\vec{\nabla}\times\vec{\mathbb{E}}=\frac{\partial\vec{\mathbb{B}}}{\partial t}+\vec{j}_{mD}\;,
\end{eqnarray}
where we use symbols ``$\mathbb{E}$'' and ``$\mathbb{B}$'' to denote the DP induced electric and magnetic fields, respectively.
After
applying the curl differential operator to the above equations, one obtains two second-order differential equations
\begin{eqnarray}
\vec{\nabla}^2\vec{\mathbb{E}}-\frac{\partial^2 \vec{\mathbb{E}}}{\partial t^2}&=&\frac{\partial \vec{j}_{eD}}{\partial t}+\vec{\nabla}\times \vec{j}_{mD}\;,
\label{eq:E2}\\
\vec{\nabla}^2\vec{\mathbb{B}}-\frac{\partial^2 \vec{\mathbb{B}}}{\partial t^2}&=&\frac{\partial \vec{j}_{mD}}{\partial t}-\vec{\nabla}\times \vec{j}_{eD}\;,
\label{eq:B2}
\end{eqnarray}
where we take $A_D^0=0$ or $\tilde{A}_D^0=\tilde{B}_D^0=0$, and only keep the spatial components of them~\cite{Arias:2012az}. Next, we consider two cases for the dark currents corresponding to the above two types of Maxwell's equations, respectively
\begin{eqnarray}
{\rm case~I}:~~\left\{
  \begin{array}{ll}
    \vec{j}_{eD}=\epsilon_1 m_D^2 \vec{A}_D\;, & \\
    \vec{j}_{mD}=\epsilon_2 m_D^2 \vec{A}_D\;, &
  \end{array}
\right.~~
{\rm case~II}:~~\left\{
  \begin{array}{ll}
    \vec{j}_{eD}=\epsilon m_D^2 \vec{\tilde{A}}_D\;, & \\
    \vec{j}_{mD}=\epsilon m_D^2 \vec{\tilde{B}}_D\;. &
  \end{array}
\right.
\end{eqnarray}
The two cases also correspond to one-component ($A_D$) DM scenario or two-component ($\tilde{A}_D$ and $\tilde{B}_D$) DM scenario.
In these two cases, the local DM density~\cite{Turner:1990qx,ADMX:2021nhd} is given by
\begin{eqnarray}
\rho_0=0.45~{\rm GeV}~{\rm cm}^{-3}=
\left\{
  \begin{array}{ll}
    {1\over 2}m_D^2|\vec{A}_D|^2 & {\rm case~I}\;,\\
    {1\over 2}m_D^2 (|\vec{\tilde{A}}_D|^2 + |\vec{\tilde{B}}_D|^2) & {\rm case~II}\;.
  \end{array}
\right.
\end{eqnarray}
We adopt the scenario in Refs.~\cite{Arias:2012az,Arias:2014ela} to ensure that the dark photon field is along a fixed direction $\vec{k}$. As a result, $\vec{\nabla}\times\vec{j}_{eD}=\vec{\nabla}\times\vec{j}_{mD}=0$.
In case II, we define the ratio of two-component DM percentages as
\begin{eqnarray}
{|\vec{\tilde{A}}_D|^2 \over |\vec{\tilde{B}}_D|^2} ={x\over 1-x}\;,
\end{eqnarray}
where $0 < x < 1$.
Then, the DP DM fields can be expressed as follows
\begin{eqnarray}
{\rm case~I}:~~\vec{A}_D={\sqrt{2\rho_0}\over m_D}e^{-im_Dt} \hat{\vec{k}}\;,~~{\rm case~II}:~~\left\{
  \begin{array}{ll}
    \vec{\tilde{A}}_D={\sqrt{2\rho_0 x}\over m_D}e^{-im_Dt} \hat{\vec{k}} & \;,\\
    \vec{\tilde{B}}_D={\sqrt{2\rho_0 (1-x)}\over m_D}e^{-im_Dt} \hat{\vec{k}} & \;.
  \end{array}
\right.
\end{eqnarray}
One can see that in case I, the DM density is composed of $A_D$ only. The two second-order differential equations are governed by two kinetic mixing parameters $\epsilon_1$ and $\epsilon_2$, respectively. In case II, there is one free kinetic mixing parameter $\epsilon$. The two equations are induced by the two components of DM $\vec{\tilde{A}}_D$ and $\vec{\tilde{B}}_D$, respectively.

Note that the Lagrangian Eq.~(\ref{eq:DP-P}) satisfies $SL(2,Z)$ symmetry which ensures the theory's consistency under electromagnetic dual transformations. The symmetry implies that electric and magnetic charges can be interchanged under $SL(2,Z)$ transformation, with the Lagrangian form remaining invariant. One can rewrite the general QEMD Lagrangian in terms of differential form notation~\cite{Csaki:2010rv}
\begin{equation}
\begin{split}
\mathcal{L}_{\rm P}&=-{\rm Im}\Big\{\frac{\tau}{8\pi n^2}[n\cdot\partial\wedge(\overline{A}+i\overline{B})]\cdot[n\cdot\partial\wedge(\overline{A}-i\overline{B})]\Big\}\\
&-{\rm Re}\Big\{\frac{\tau}{8\pi n^2}[n\cdot\partial\wedge(\overline{A}+i\overline{B})]\cdot[n\cdot(\partial\wedge(\overline{A}-i\overline{B}))^d]\Big\}
-{\rm Re}\Big\{(\overline{A}-i\overline{B})\cdot(J+\tau K)\Big\}\;,
\end{split}
\label{equ:Zwanziger}
\end{equation}
where $\tau$ is the modular group parameter of the $SL(2,Z)$ symmetry
\begin{eqnarray}
\tau={\theta\over 2\pi}+{in_0\over e^2}
\end{eqnarray}
with $n_0=eg=4\pi$, and we neglect $\theta$ below. We redefine $\overline{A}\equiv eA$, $\overline{B}\equiv eB$, $J\equiv j_e/e$ and $K\equiv j_m e/4\pi$.
Under the $SL(2,Z)$ duality transformation, the parameter $\tau$ and the electromagnetic fields are transformed as~\cite{Terning:2018lsv}
\begin{eqnarray}
&&\tau\to \tau'= {a\tau+b\over c\tau +d}\;,~~~{\rm Im}(\tau)\to {\rm Im}(\tau')={{\rm Im}(\tau)\over |c\tau +d|^2}\;,\\
&&\overline{A}_\mu+i\overline{B}_\mu\to\frac{1}{c\tau^*+d}(\overline{A}'_\mu+i\overline{B}'_\mu)\;,\\
&&\overline{A}_\mu-i\overline{B}_\mu\to\frac{1}{c\tau+d}(\overline{A}'_\mu-i\overline{B}'_\mu)\;,
\end{eqnarray}
where the integers $a$, $b$, $c$ and $d$ are the matrix elements of $SL(2,Z)$ transformation and satisfy $ad-bc=1$.
The electromagnetic tensors and currents follow the transformations~\cite{Terning:2018lsv}
\begin{eqnarray}
&&\overline{F}_{\mu\nu}+i\overline{F}_{\mu\nu}^d\to\frac{1}{c\tau^*+d}(\overline{F}'_{\mu\nu}+i\overline{F}_{\mu\nu}^{\prime~d})\;,\\
&&\overline{F}_{\mu\nu}-i\overline{F}_{\mu\nu}^d\to\frac{1}{c\tau+d}(\overline{F}'_{\mu\nu}-i\overline{F}_{\mu\nu}^{\prime~d})\;,\\
&&J\to bK'+dJ'\;,~~~K\to aK'+cJ'\;.
\end{eqnarray}
We can also rewrite our Lagrangian in Eq.~(\ref{eq:DP-P}) as
\begin{eqnarray}
\mathcal{L}_{\rm DP-P}&=&-\frac{{\rm Im}(\tau)}{4\pi}\Big\{\frac{\epsilon_1}{2}\left[\overline{F}\cdot(\partial\wedge \overline{A}_D)+\overline{F}^d\cdot(\partial\wedge \overline{B}_D)\right]\nonumber\\
&&+\frac{\epsilon_2}{2}\left[\overline{F}^d\cdot(\partial\wedge \overline{A}_D)+\overline{F}\cdot(\partial\wedge \overline{B}_D)\right]\Big\}\;.
\end{eqnarray}
The complete equation of motion is
\begin{eqnarray}
{{\rm Im}(\tau)\over 4\pi} \partial_\mu (\overline{F}+i\overline{F}^d)- {{\rm Im}(\tau)\over 4\pi} m_D^2 (\epsilon_1+i\epsilon_2)(\overline{A}_D+i\overline{B}_D) = J + \tau K\;.
\end{eqnarray}
Under $SL(2,Z)$, it then becomes
\begin{eqnarray}
{{\rm Im}(\tau')\over 4\pi} \partial_\mu (\overline{F}'+i\overline{F}^{\prime d})- {{\rm Im}(\tau')\over 4\pi} m_D^2 (\epsilon_1+i\epsilon_2)(\overline{A}^\prime_D+i\overline{B}^\prime_D) = J' + \tau' K'\;.
\end{eqnarray}

\section{Strategy and sensitivity of haloscope experiments}
\label{sec:Sens}

Next, we solve the above equations in terms of the DP DM fields, and examine the detection strategies in cavity haloscope experiment~\cite{Arias:2012az,Yang:2022uil} or LC circuit experiment~\cite{Arias:2014ela,Chaudhuri:2014dla}~\footnote{There is also detection strategy of vector DM using the Zeeman effect between atomic states~\cite{Yang:2016zaz}.}. Below we take case II as an illustrative investigation and solve the two equations governed by $\vec{\tilde{A}}_D$ and $\vec{\tilde{B}}_D$, respectively. The results of case I can be easily obtained by taking the substitution $\epsilon\sqrt{x}\to \epsilon_1$ or $\epsilon\sqrt{1-x}\to \epsilon_2$.

For the direction of DP DM, we assume $\theta$ as the angle between the direction $\vec{k}$ of the DP field and the $z$ direction in the laboratory coordinate system~\cite{Arias:2012az}. The direction $\vec{k}$ of the DP field can be arbitrary. We need to average the final result over all randomly pointing directions for $\vec{k}$.

\subsection{cavity haloscope}

We first revisit the solution of Eq.~(\ref{eq:E2}) in case II for cavity experiment. It is exactly the same Maxwell's equation induced by DP DM electrodynamics for conventional cavity experiment.

The electric field $\vec{\mathbb{E}}(t,\vec{x})$ in microwave cavity can be decomposed as the superposition of the time-evolution functions $e_n(t)$ and orthogonal modes $\mathbb{E}_n(\vec{x})$
\begin{eqnarray}
\vec{\mathbb{E}}(t,\vec{x})=\sum_n e_n(t) \vec{\mathbb{E}}_n(\vec{x})\;,
\label{equ:superposition of E}
\end{eqnarray}
where modes $\vec{\mathbb{E}}_n(\vec{x})$ satisfy the Helmholtz equation $\vec{\nabla}^2\vec{\mathbb{E}}_n+\omega_n^2\vec{\mathbb{E}}_n=0$ with the resonant frequency $\omega_n$ being equal to the frequency of DP $\omega_D\approx m_D$. Plugging $\vec{\mathbb{E}}(t,\vec{x})$ into Eq.~(\ref{eq:E2}) and considering the losses within cavity, we obtain the expansion coefficient $e_n(t)$ as follows
\begin{eqnarray}
\left(\frac{d^2}{dt^2}+\frac{\omega_D}{Q}\frac{d}{dt}+\omega_D^2\right)e_n(t)=-\frac{\epsilon m_D^2}{C_n^\mathbb{E}}\int{dV \vec{\mathbb{E}}_n^*(\vec{x})\cdot \partial_t\vec{\tilde{A}}_D}\;,
\label{eq:coeffient equation}
\end{eqnarray}
where the normalization coefficients are defined as $C_n^\mathbb{E}=\int{dV\vert \vec{\mathbb{E}}_n(\vec{x})\vert^2}$ and $Q$ denotes quality factor. When assuming $e_n(t)=e_{n,0}e^{-i\omega t}$, the coefficient $e_{n,0}$ is given by
\begin{eqnarray}
e_{n,0}&=&\frac{(\omega_D^2-\omega^2+i\frac{\omega\omega_D}{Q})}{(\omega_D^2-\omega^2)^2+\frac{\omega^2\omega_D^2}{Q^2}}\bigg\vert_{\omega\approx\omega_D}\times\left(-\frac{\epsilon m_D^2}{C_n^\mathbb{E}}\right)\int{dV \vec{\mathbb{E}}_n^*\cdot \partial_t\vec{\tilde{A}}_D}\nonumber\\
&=&-i\frac{\epsilon Q}{C_n^\mathbb{E}}\times\int{dV \vec{\mathbb{E}}_n^*(\vec{x})\cdot \partial_t\vec{\tilde{A}}_D}\;.
\end{eqnarray}
The output power in the cavity can be obtained in terms of the energy stored in the cavity $U$ and the quality factor
\begin{eqnarray}
P_{\rm DP}^\mathbb{E}=\kappa \frac{U}{Q}\omega_D=\kappa\frac{\omega_D}{Q}\frac{|e_{n,0}|^2}{2}\int{dV|\vec{\mathbb{E}}_n(\vec{x})|^2}=\frac{\kappa}{2}\epsilon^2 m_D QV\vert\partial_t \vec{\tilde{A}}_D\vert^2 G^\mathbb{E}\cos^2\theta\;,
\label{eq:Epower}
\end{eqnarray}
where $\kappa$ is the cavity coupling factor depending on the experimental setup, $\vert\partial_t \vec{\tilde{A}}_D\vert^2=2\rho_0 x$, and the form factor depending on the geometry of cavity is
\begin{eqnarray}
G^\mathbb{E}=\frac{\vert\int{dV \vec{\mathbb{E}}_n^*(\vec{x})\cdot \vec{z}} \vert^2}{V\int{dV \vert \vec{\mathbb{E}}_n(\vec{x})\vert^2}} \;.
\end{eqnarray}
After averaging over all possible directions of DP, compared to the axion cavity detection, the form factor here should be multiplied by $\langle\cos^2\theta\rangle=\int{\cos^2\theta d\Omega}/\int{d\Omega}=1/3$.

\begin{figure}[tb!]
\centering
\includegraphics[width=0.7\textwidth]{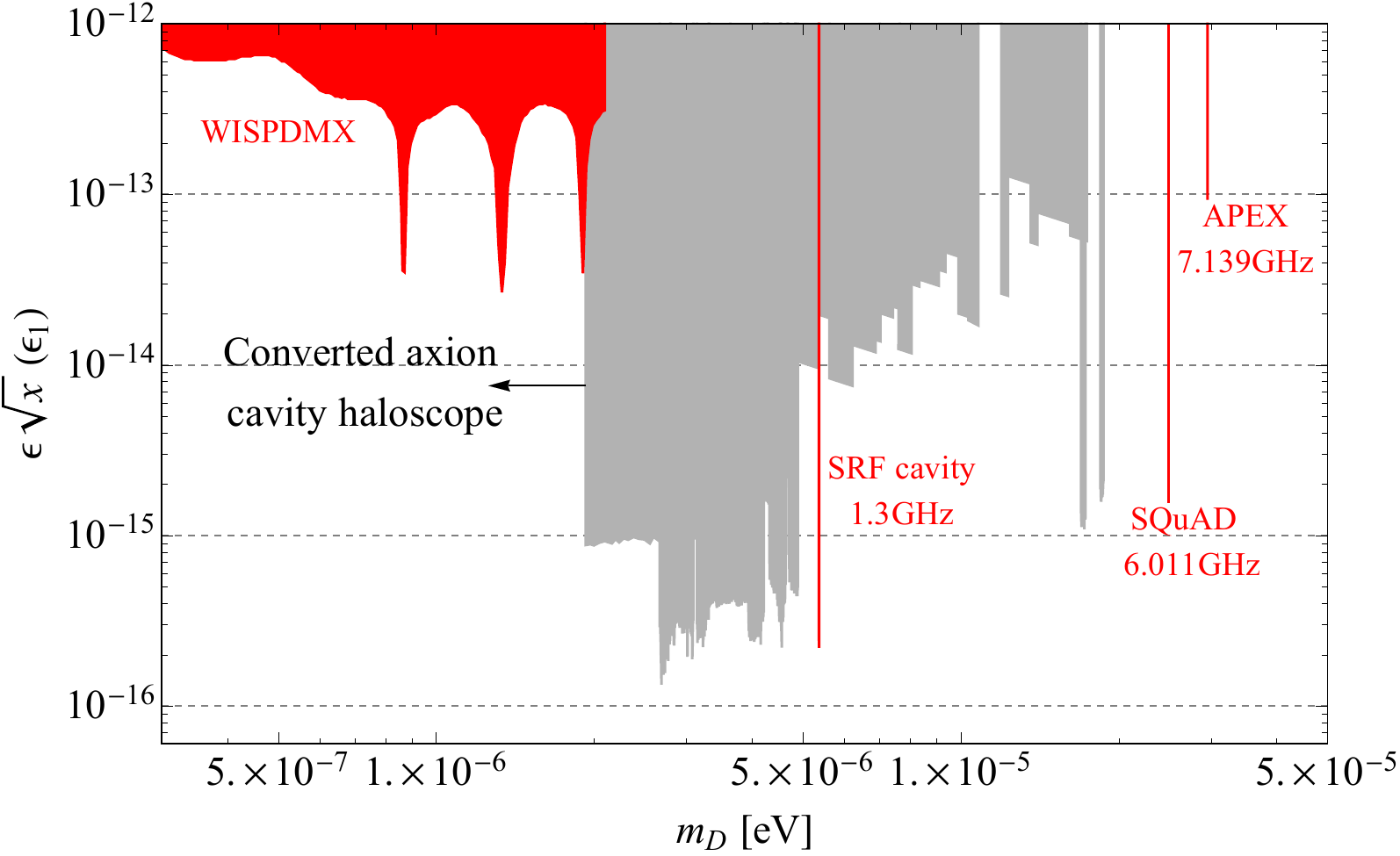}
\caption{The sensitivity of cavity haloscope experiment to kinetic mixing $\epsilon \sqrt{x}~(\epsilon_1)$. For the converted results in gray, we take the limits of axion coupling $g_{a\gamma\gamma}$ from the AxionLimits repository~\cite{AxionLimits}. The direct DP search limits from WISPDMX~\cite{Nguyen:2019xuh}, SRF cavity~\cite{Tang:2023oid}, SQuAD~\cite{Dixit:2020ymh} and APEX~\cite{He:2024ytp} are also shown in red.
}
\label{fig:cavity}
\end{figure}

It was demonstrated that for
the detection of the DP field, similar to the axion search in cavity experiments, the ${\rm TM}_{010}$ mode has the largest coupling to DP with the electric field along the $\vec{z}$-axis.
In this case, the theoretical value of the form factor for ideal cylindrical cavity is $G^\mathbb{E}\approx 0.69$ and the value for ADMX with tuning rods is $G^\mathbb{E}\approx 0.455$~\cite{ADMX:2021nhd}.
The signal power in axion cavity experiments can be shown as
\begin{eqnarray}
P_{\rm axion}^\mathbb{E}=\kappa \left[g_{a\gamma\gamma}^2\frac{|\mathbb{B}_0|^2}{m_a}\right]\rho_0 QVG^\mathbb{E}\;,
\end{eqnarray}
where $|\mathbb{B}_0|$ denotes the magnitude of the external static magnetic field. Taking $P_{\rm DP}^\mathbb{E}=P_{\rm axion}^\mathbb{E}$ yields the relation
\begin{eqnarray}
\epsilon^2 m_D x \langle\cos^2\theta\rangle = g_{a\gamma\gamma}^2 \frac{|\mathbb{B}_0|^2}{m_a} \;.
\label{eq:relation}
\end{eqnarray}
The constraints on the axion-photon coupling $g_{a\gamma\gamma}$ from the existing axion cavity experiments can be converted to constrain the parameter $\epsilon \sqrt{x}$ for DP. Fig.~\ref{fig:cavity} shows the sensitivity of cavity haloscope experiment to kinetic mixing $\epsilon \sqrt{x}$ (or $\epsilon_1)$ as a function of $m_D$. For $\epsilon \sqrt{x}$, we show the converted limits (gray) from axion cavity experiments according to Eq.~(\ref{eq:relation}) as well as direct DP search limits (red) from WISPDMX~\cite{Nguyen:2019xuh}, SRF cavity~\cite{Tang:2023oid}, SQuAD~\cite{Dixit:2020ymh} and APEX~\cite{He:2024ytp}.
It is clear that the form of our Maxwell's equation Eq.~(\ref{eq:E2}) is exactly the same as the one for DP in QED. The induced power in cavity in Eq.~(\ref{eq:Epower}) is also similar to that in QED~\cite{Arias:2012az}, except for the additional free DM fraction constant $x$. This is because we have two-component DP DM here in the QEMD model ($\tilde{A}_D$ and $\tilde{B}_D$), rather than the QED case in which one-component DP comprises 100\% DM density. In other words, when $x=1$, our analytical results of $\tilde{A}_D$ will restore the QED DP case. The spirit of experimental setup for $\tilde{A}_D$ should also be the same as that for QED DP. Thus, the existing results of current DP cavity experiments can be directly applied to constrain our parameter combination $\epsilon\sqrt{x}$ in case II (or $\epsilon_1$ in our case I). The above WISPDMX, SRF cavity, SQuAD and APEX experiments searched for resonant DPs using a tunable radio frequency cavity or a superconducting radio frequency cavity with a high quality factor. They provide upper bounds on $\epsilon\sqrt{x}$ for individual DP masses in the range of $\sim 0.1 - 30~\mu{\rm eV}$. The SRF cavity with a remarkably high quality factor of about $10^{10}$ yields the most stringent bound as $\epsilon\sqrt{x}<2\times 10^{-16}$ at 1.3 GHz~\cite{Tang:2023oid}. APEX uses high-performance amplifiers specifically designed for low-temperature environments, achieving an extremely low temperature compared to other experiments and effectively reducing background noise. It sets the parameter limit as $\epsilon\sqrt{x}<3.7\times 10^{-13}$ around 7.139 GHz at 90\% confidence level~\cite{He:2024ytp}. SQuAD employs superconducting qubit technology and sub-standard quantum limit (sub-SQL) detection techniques, further reducing noise and improving detection precision. Its background shot noise remains at 15.7 dB, providing a limit of $\epsilon\sqrt{x}<1.68\times10^{-15}$ at 6.011 GHz~\cite{Dixit:2020ymh}. WISPDMX employs four tuned resonant modes to scan for signals, enabling a probe over a broader mass range of $0.2–2.07~\mu{\rm eV}$ and achieving a sensitivity limit of $\epsilon\sqrt{x}<10^{-13}-10^{-12}$~\cite{Nguyen:2019xuh}.

Similarly, we can follow the same procedure to obtain the solution of Eq.~(\ref{eq:B2}) for the emission power of magnetic field modes $\vec{\mathbb{B}}_n(\vec{x})$ induced by $\partial_t \vec{\tilde{B}}_D$
\begin{eqnarray}
P_{\rm DP}^\mathbb{B}&=&\frac{\kappa}{2}\epsilon^2m_DQV\vert\partial_t \vec{\tilde{B}}_D \vert^2 G^\mathbb{B}\cos^2\theta = \kappa \epsilon^2m_D (1-x)\rho_0 QV G^\mathbb{B}\cos^2\theta,\\
G^\mathbb{B}&=&\frac{\vert\int{dV \vec{\mathbb{B}}_n^*(\vec{x})\cdot \vec{z}} \vert^2}{V\int{dV \vert \vec{\mathbb{B}}_n(\vec{x})\vert^2}}\;.
\label{equ:Bpower}
\end{eqnarray}
Next, we discuss the feasibility of DP field $\vec{\tilde{B}}_D$ detection.
It turns out that one needs the magnetic field modes along the $\vec{z}$-axis, corresponding to the TE modes.
In an ideal cylindrical cavity, the form factor of ${\rm TE}_{011}$ mode is canceled over the radial integral from 0 to the radius. The higher-order ${\rm TE}_{111}$ mode exhibits periodic symmetry along the direction of azimuthal angle $\phi$. This causes the response to dark photon to be canceled over the integral from 0 to $2\pi$. To avoid the cancellation, taking ${\rm TE}_{111}$ mode for illustration, we need imperfectly symmetric field distribution over the azimuthal angle direction within the cavity. In the practical setup of cavity experiment, tuning rods are placed within the cylindrical cavity in order to tune the mode. For instance, in the ADMX experiment, two copper rods are put inside the cavity for frequency tuning. As indicated in Ref.~\cite{ADMX:2019uok}, the tuning rods rotate around a fixed center and then the field distribution in cavity
transforms asymmetrically over a full scan cycle. Here, a reasonable modification for the ${\rm TE}_{111}$ mode is to take only one half of the original cylindrical cavity. We then partially integrate the magnetic field and calculate the form factors. Specifically, after integrating the azimuthal angle from 0 to $\pi$, the analytical result of the form factor $G^\mathbb{B}$ corresponding to ${\rm TE}_{111}$ mode is $(128/\pi^4 x_{1,1}^{\prime2})\times(c_1^2/c_2)\approx 0.61$, where $c_1=\int_0^{x_{1,1}^\prime}{dx x J_1(x)}$ and $c_2=\int_0^{x_{1,1}^\prime}{dx x J_1^2(x)}$ with $x^\prime_{m,n}$ being the $n$-th zero point of the first derivative of Bessel function $J_m(x)$. It is close to the form factor of ${\rm TM}_{010}$ mode. We leave the detailed electromagnetic simulation in a future work.

\subsection{LC circuit}

For LC circuit experiment, one needs to solve the DP Maxwell's equations with electromagnetic shielding~\cite{Chaudhuri:2014dla}. We take the shield as a conducting or superconducting hollow cylinder of radius $R$ along the $\hat{z}$ direction in cylindrical coordinates $(\rho,\phi,z)$. In our case, in the presence of $\vec{j}_{eD}$ and $\vec{j}_{mD}$, both the induced electric field and magnetic field in $z$ direction would respectively be suppressed as a result of the electromagnetic shielding. That is to say the observed $\vec{\mathbb{E}}$ and $\vec{\mathbb{B}}$ fields should be solved under the boundary conditions $\hat{z}\cdot \vec{\mathbb{E}}=\hat{z}\cdot \vec{\mathbb{B}}=0$ on the surface with $\rho=R$~\cite{Chaudhuri:2014dla}. The $\vec{\mathbb{B}}$ and $\vec{\mathbb{E}}$ along the $\phi$ direction generated by the currents then become the dominant observable fields inside the shield.

The DP field is projected to the $z$ direction below, and thus we have $\vec{j}_{eD}=\epsilon m_D \sqrt{2\rho_0 x} e^{-im_D t}\cos\theta \hat{z}$ and $\vec{j}_{mD}=\epsilon m_D \sqrt{2\rho_0 (1-x)} e^{-im_D t}\cos\theta \hat{z}$.
For the current $\vec{j}_{eD}$ induced by $\vec{\tilde{A}}_{D}$, we solve the following equations
\begin{eqnarray}
\vec{\nabla}^2\vec{\mathbb{E}}-\frac{\partial^2 \vec{\mathbb{E}}}{\partial t^2}&=&\frac{\partial \vec{j}_{eD}}{\partial t}\;,\\
\vec{\nabla}\times\vec{\mathbb{B}}&=&\frac{\partial\vec{\mathbb{E}}}{\partial t}+\vec{j}_{eD}\;.
\end{eqnarray}
The solution of observable $\vec{\mathbb{E}}$ and $\vec{\mathbb{B}}$ becomes
\begin{eqnarray}
\vec{\mathbb{E}}_{\rm obs}&=&-i\epsilon \sqrt{2\rho_0 x}\cos\theta e^{-im_D t} \Big(1- {J_0(m_D\rho)\over J_0(m_D R)} \Big)\hat{z}\nonumber \\
&\approx& i\epsilon \sqrt{2\rho_0 x}\cos\theta e^{-im_D t} m_D^2 (R^2-\rho^2) \hat{z}\;,
\label{eq:Esol}\\
\vec{\mathbb{B}}_{\rm obs}&=&\epsilon \sqrt{2\rho_0 x}\cos\theta e^{-im_D t} {J_1(m_D\rho)\over J_0(m_D R)} \hat{\phi}\approx \epsilon \sqrt{2\rho_0 x}\cos\theta e^{-im_D t} m_D \rho \hat{\phi}\;.
\label{eq:Bsol}
\end{eqnarray}
When $x\to 1$, it is exactly the same as the solution in Ref.~\cite{Chaudhuri:2014dla}. An ajustable LC circuit is put inside a hollow conducting shield, and the inducting coil is wrapped around the $\phi$ direction of the conductor to receive the
driving magnetic field.
When the resonant frequency of the LC circuit is tuned to the DP oscillation frequency, the observable magnetic field gives rise to the magnetic flux and the consequent current
\begin{eqnarray}
\Phi_{\rm obs}\approx Q \epsilon \sqrt{2\rho_0 x}\cos\theta m_D V \;,~~I={\Phi_{\rm obs}\over L}\;,
\end{eqnarray}
where $Q\sim 10^6$ is the quality factor of the LC circuit, $V$ is the volume of the inductor and $L$ is the inductance of the inducting coil. The signal power is then given by
\begin{eqnarray}
P_{\rm signal}= \langle I^2 R_s \rangle
\approx {\rho_0 x Q\epsilon^2 \langle\cos^2\theta\rangle m_D^3 V^2\over L}
\approx \rho_0 x Q\epsilon^2 \langle\cos^2\theta\rangle m_D^3 V^{5/3}\;,
\end{eqnarray}
where $R_s=Lm_D/Q$ is the resistance.
The solutions of electromagnetic fields in Eq.~(\ref{eq:Esol}) and Eq.~(\ref{eq:Bsol}) are analogous to those of DP in QED case~\cite{Chaudhuri:2014dla}, except for the DM fraction constant $x$. The signal power is also similar by multiplying an additional $x$ factor. One can thus arrange the same setup of LC circuit experiments here for $\tilde{A}_D$. The existing limits of DP from QED experiments can be directly applied to constrain the parameter combination $\epsilon\sqrt{x}$ in case II (or $\epsilon_1$ in case I) of our QEMD model.

The equations for current $\vec{j}_{mD}$ induced by $\vec{\tilde{B}}_{D}$ are
\begin{eqnarray}
\vec{\nabla}^2\vec{\mathbb{B}}-\frac{\partial^2 \vec{\mathbb{B}}}{\partial t^2}&=&\frac{\partial \vec{j}_{mD}}{\partial t}\;,\\
-\vec{\nabla}\times\vec{\mathbb{E}}&=&\frac{\partial\vec{\mathbb{B}}}{\partial t}+\vec{j}_{mD}\;.
\end{eqnarray}
The solution is
\begin{eqnarray}
\vec{\mathbb{B}}_{\rm obs}&=&-i\epsilon \sqrt{2\rho_0 (1-x)}\cos\theta e^{-im_D t} \Big(1- {J_0(m_D\rho)\over J_0(m_D R)} \Big)\hat{z}\nonumber \\
&\approx& i\epsilon \sqrt{2\rho_0 (1-x)}\cos\theta e^{-im_D t} m_D^2 (R^2-\rho^2) \hat{z}\;,\\
\vec{\mathbb{E}}_{\rm obs}&=&-\epsilon \sqrt{2\rho_0 (1-x)}\cos\theta e^{-im_D t} {J_1(m_D\rho)\over J_0(m_D R)} \hat{\phi}\approx -\epsilon \sqrt{2\rho_0 (1-x)}\cos\theta e^{-im_D t} m_D \rho \hat{\phi}\;.
\end{eqnarray}
In this case, a superconducting shield is placed outside the electromagnetic detector. The magnetic field in $z$ direction is suppressed due to the superconducting Meissner effect.
A wire loop is put inside the cylindrical hole of the superconducting shield to conduct the induction current~\cite{Li:2022oel}. The LC circuit is then connected to the wire loop to enhance the signal power. The induction current is
\begin{eqnarray}
I={2\pi R \mathbb{E}_{\rm obs}(R)\over R_s}={2\pi R^2 \epsilon \sqrt{2\rho_0 (1-x)}\cos\theta m_D \over R_s}\;.
\end{eqnarray}
The signal power is then given by
\begin{eqnarray}
P_{\rm signal}=\langle I^2 R_s \rangle
\approx {\rho_0(1-x)Q\epsilon^2\langle\cos^2\theta\rangle m_DV^{4/3}\over L}\approx \rho_0(1-x)Q\epsilon^2\langle\cos^2\theta\rangle m_D V\;.
\end{eqnarray}

\begin{figure}[tb!]
\centering
\includegraphics[width=0.7\textwidth]{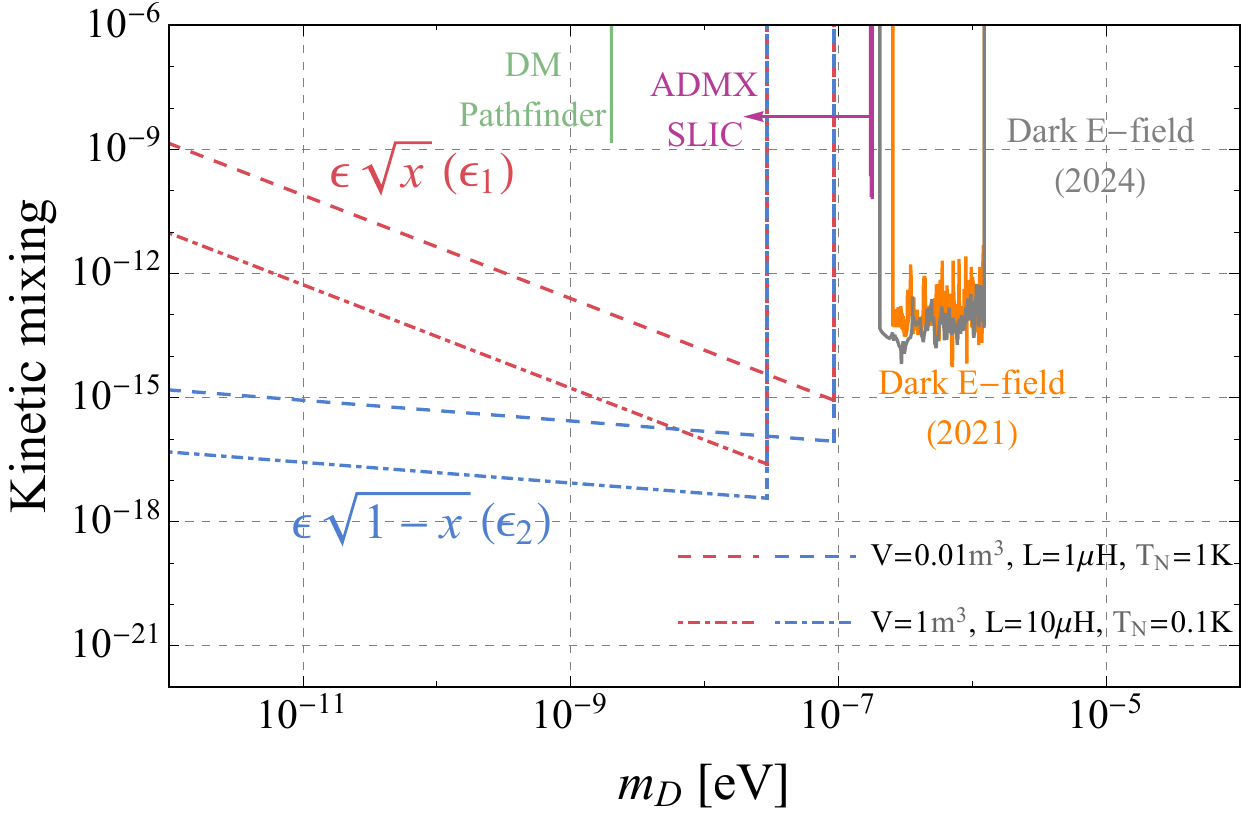}
\caption{The sensitivity of LC circuit to kinetic mixings $\epsilon \sqrt{x}~(\epsilon_1)$ (red) and $\epsilon \sqrt{1-x}~(\epsilon_2)$ (blue). We assume two setup benchmarks for both cases. Some existing limits for light DP DM are also shown, including DM Pathfinder (green)~\cite{Phipps:2019cqy}, ADMX SLIC (purple)~\cite{Crisosto:2019fcj}, Dark E-Field Radio Experiment (orange)~\cite{Godfrey:2021tvs} and (gray)~\cite{Levine:2024noa}.
}
\label{fig:LC}
\end{figure}

We adopt the cryogenic amplifier described in Ref.~\cite{Duan:2022nuy} to receive
and amplify the signals.
The thermal noise exists in circuit can be estimated as
\begin{eqnarray}
P_{\rm noise}=\kappa_B T_N \sqrt{\Delta f\over \Delta t}\;,
\end{eqnarray}
where $\kappa_B$ is the Boltzmann constant, $T_N$ is the noise temperature, $\Delta f=f/Q$ is the detector bandwidth and $\Delta t$ is the observation time. We take one week of observation time and two setup benchmarks of volume $V$, inductance $L$ and temperature $T_N$ for both cases. An adjustable capacitance with a minimal value of 50 pF is taken, resulting a
maximal frequency.
To estimate the sensitivity of $\epsilon \sqrt{x}$ or $\epsilon \sqrt{1-x}$, we require the signal-to-noise ratio (SNR) to satisfy
\begin{eqnarray}
{\rm SNR}={P_{\rm signal}\over P_{\rm noise}}>3\;.
\end{eqnarray}
In Fig.~\ref{fig:LC}, we show the sensitivity of LC circuit to $\epsilon\sqrt{x}~(\epsilon_1)$ (red) and $\epsilon\sqrt{1-x}~(\epsilon_2)$ (blue). The search potential of $\vec{\tilde{B}}_D$ is more promising than that of $\vec{\tilde{A}}_D$ at low frequencies. Some exclusion limits for light DP DM are also shown, including DM Pathfinder (green)~\cite{Phipps:2019cqy}, ADMX SLIC (purple)~\cite{Crisosto:2019fcj}, Dark E-Field Radio Experiment (orange)~\cite{Godfrey:2021tvs} and (gray)~\cite{Levine:2024noa}. As stated before, they can be applied to constrain our parameter combination $\epsilon\sqrt{x}$ in case II (or $\epsilon_1$ in case I). These experiments are sensitive to DP mass lower than about $1~\mu{\rm eV}$. An early fixed-frequency superconducting resonator sets a simple exclusion limit on $\epsilon\sqrt{x}>1.5\times 10^{-9}$ for $\sim 2~{\rm neV}$ DPs~\cite{Phipps:2019cqy}. The most recent Dark E-Field Radio experiment can place a 95\% exclusion limit on $\epsilon\sqrt{x}$ between $6\times 10^{-15}$ and $6\times 10^{-13}$ over the mass range of $0.21 – 1.24~\mu{\rm eV}$~\cite{Levine:2024noa}. The ADMX SLIC experiment uses a superconducting LC circuit to detect low-frequency light axions in strong magnetic fields (ranging from 4.5 T to 7.0 T). When rescaled for our DP-photon kinetic mixing parameter, its exclusion limit gives $\epsilon\sqrt{x}<10^{-10}$ in the sub $0.2~\mu{\rm eV}$ range~\cite{Crisosto:2019fcj}.

\subsection{Connection to cosmology}

Next, we briefly discuss the connection of DP DM to cosmology. There are a few plausible DP production mechanisms that may generate the correct abundance of DM in the early Universe. The most popular one is the misalignment mechanism~\cite{Preskill:1982cy,Abbott:1982af,Dine:1982ah}. Ref.~\cite{Nelson:2011sf} verified that the misalignment mechanism for axion also applied to DP. Next, we will explore the evolution equation and energy density of QEMD DP in a cosmological context. For simplicity, we focus on the homogeneous solution of QEMD DP fields with $\partial_i {A_D}_\mu=\partial_i {B_D}_\mu=0$. In an expanding universe, we adopt the metric as ${\rm diag}(1, -a^2, -a^2,-a^2)$ with $a(t)$ being scale factor. We define the following anti-symmetric tensors
\begin{eqnarray}
&&G_{\alpha\beta}={F_D}_{\alpha\beta}+\epsilon_1 F_{\alpha\beta}+\epsilon_2 F^d_{\alpha\beta}\;,~~K_{\alpha\beta}=\epsilon_1 {F_D}_{\alpha\beta}-\epsilon_2 {F_D^d}_{\alpha\beta}\;,\\
&&\bar{G}_{\alpha\beta}=G_{\alpha\beta}/a^2(t)\;,~~\bar{K}_{\alpha\beta}=K_{\alpha\beta}/a^2(t)\;,
\end{eqnarray}
where $\epsilon_1$ and $\epsilon_2$ are the two DP-photon mixing couplings. Under these conventions, we can write down the evolution equations for the QEMD DP fields in the Universe as
\begin{eqnarray}
\partial_0G_{0\beta}+3HG_{0\beta}-\partial_i {\bar{G}_{i\beta}}+m_D^2 {A_D}_\beta&=0\;,\\
\partial_0G_{0\beta}^d+3HG_{0\beta}^d-\partial_i {\bar{G}_{i\beta}^d}+m_D^2 {B_D}_\beta&=0\;,
\end{eqnarray}
where $H$ denotes the Hubble parameter and we neglect the effect of non-minimal coupling(s) to gravity without changing the conclusion~\cite{Arias:2012az}.
The corresponding energy density is given by
\begin{eqnarray}
\rho(t)&=&T^0_{~0}\nonumber\\
&=&-(\epsilon_1F_D^{0i}-\epsilon_2 {F_D^d}^{0i})(\partial_0 A_i)-(F_D^{0i}+\epsilon_1F^{0i}+\epsilon_2 {F^d}^{0i})(\partial_0 {A_D}_i)\nonumber\\
&&-(\epsilon_1 {F_D^d}^{0i}+\epsilon_2 F_D^{0i})(\partial_0 B_i)-({F_D^d}^{0i}+\epsilon_1 {F^d}^{0i}-\epsilon_2 F^{0i})(\partial_0 {B_D}_i)-\mathcal{L}_{\rm DP}-\mathcal{L}_{\rm DP-P}\nonumber\\
&=&-K^{0i}(\partial_0A_i)-G^{0i}(\partial_0 {A_D}_i)-{K^d}^{0i}(\partial_0 B_i)-{G^d}^{0i}(\partial_0{B_D}_i)-\mathcal{L}_{\rm DP}-\mathcal{L}_{\rm DP-P}\nonumber\\
&=&\bar{K}_{0i}\dot{A}_i+\bar{G}_{0i}\dot{A}_{Di}+\bar{K}_{0i}^d\dot{B}_i+\bar{G}_{0i}^d\dot{B}_{Di}-\mathcal{L}_{\rm DP}-\mathcal{L}_{\rm DP-P}\;.
\end{eqnarray}
It is easy to prove that the above results can reduce to that of QED DP when $B$ and $B_D$ fields vanish. This consistency indicates that the QEMD DP model is a reasonable extension of QED DP in cosmology. As a result, the energy density of DP behaves as non-relativistic matter
with $\rho(t)\propto 1/a^3(t)$~\cite{Arias:2012az}.

Another scenario of DP DM production is through quantum fluctuations during inflation~\cite{Graham:2015rva}. Additionally, DP DM can be produced from the decay of topological defects such as cosmic strings~\cite{Long:2019lwl}.
This work focuses on the low-energy dynamics of DP and the detection in laboratory. One assumes that the DM distribution around the Earth is comprised of a cold population of DPs. A very detailed cosmological study of the QEMD DP is beyond the scope of this work, and we leave a dedicated study for the future.

\section{Conclusions}
\label{sec:Con}

Dark matter and magnetic monopole are two of
longstanding candidates of new physics beyond the SM. The ultralight dark photon is an intriguing bosonic dark matter. The interaction between the visible photon and dark photon is introduced by the gauge kinetic mixing between the field strength tensors of SM electromagnetic gauge group and dark Abelian gauge group. On the other hand, the relativistic electrodynamics was generalized to quantum electromagnetodynamics in the presence of both electric and magnetic charges. In QEMD theory, the physical photon is described by two four-potentials $A_\mu$ and $B_\mu$ corresponding to two $U(1)$ gauge groups $U(1)_A\times U(1)_B$.

In this work, we construct the low-energy dark photon-photon interactions in the framework of QEMD. We introduce
new heavy fermions charged under $U(1)_A\times U(1)_B$ in visible sector and $U(1)_{A_D}\times U(1)_{B_D}$ in dark sector. After integrating out the new fermions in vacuum polarization diagrams, the new dark photon-photon kinetic mixing interactions can be obtained. We derive the consequent field equations and the new Maxwell's equations in this framework. We also investigate the detection strategies of light dark photon DM as well as the generic kinetic mixings in cavity haloscope experiments and LC circuit experiments.

Finally, we give a detailed comparison between the DPs in QEMD and those in QED. Unlike QED, where each gauge field corresponds to either a visible photon or a DP, the QEMD framework introduces two Abelian gauge fields in both the visible and dark sectors. Consequently, the low-energy DP-photon Lagrangian in QEMD includes two kinetic mixing interactions, as opposed to the single interaction in conventional DP theory. The presence of two gauge fields for DPs also enables a two-component dark matter scenario, characterized by $\tilde{A}_D$ and $\tilde{B}_D$, with a single kinetic mixing parameter but different DM fraction constants.
From the solutions of the new DP Maxwell's equations, one of the two DM components ($\tilde{A}_D$) resembles the QED DP but includes an additional free DM fraction constant $x$. The other DP component ($\tilde{B}_D$) is essentially new, and its detection requires entirely new strategies in cavity and LC circuit experiments as we proposed in this article.

\acknowledgments
We would like to thank Yu Gao for useful discussions. T.~L. is supported by the National Natural Science Foundation of China (Grant No. 12375096, 12035008, 11975129).

\bibliography{refs}

\begin{thebibliography}{67}%
\makeatletter
\providecommand \@ifxundefined [1]{%
 \@ifx{#1\undefined}
}%
\providecommand \@ifnum [1]{%
 \ifnum #1\expandafter \@firstoftwo
 \else \expandafter \@secondoftwo
 \fi
}%
\providecommand \@ifx [1]{%
 \ifx #1\expandafter \@firstoftwo
 \else \expandafter \@secondoftwo
 \fi
}%
\providecommand \natexlab [1]{#1}%
\providecommand \enquote  [1]{``#1''}%
\providecommand \bibnamefont  [1]{#1}%
\providecommand \bibfnamefont [1]{#1}%
\providecommand \citenamefont [1]{#1}%
\providecommand \href@noop [0]{\@secondoftwo}%
\providecommand \href [0]{\begingroup \@sanitize@url \@href}%
\providecommand \@href[1]{\@@startlink{#1}\@@href}%
\providecommand \@@href[1]{\endgroup#1\@@endlink}%
\providecommand \@sanitize@url [0]{\catcode `\\12\catcode `\$12\catcode
  `\&12\catcode `\#12\catcode `\^12\catcode `\_12\catcode `\%12\relax}%
\providecommand \@@startlink[1]{}%
\providecommand \@@endlink[0]{}%
\providecommand \url  [0]{\begingroup\@sanitize@url \@url }%
\providecommand \@url [1]{\endgroup\@href {#1}{\urlprefix }}%
\providecommand \urlprefix  [0]{URL }%
\providecommand \Eprint [0]{\href }%
\providecommand \doibase [0]{http://dx.doi.org/}%
\providecommand \selectlanguage [0]{\@gobble}%
\providecommand \bibinfo  [0]{\@secondoftwo}%
\providecommand \bibfield  [0]{\@secondoftwo}%
\providecommand \translation [1]{[#1]}%
\providecommand \BibitemOpen [0]{}%
\providecommand \bibitemStop [0]{}%
\providecommand \bibitemNoStop [0]{.\EOS\space}%
\providecommand \EOS [0]{\spacefactor3000\relax}%
\providecommand \BibitemShut  [1]{\csname bibitem#1\endcsname}%
\let\auto@bib@innerbib\@empty
\bibitem [{\citenamefont {Holdom}(1986{\natexlab{a}})}]{Holdom:1985ag}%
  \BibitemOpen
  \bibfield  {author} {\bibinfo {author} {\bibfnamefont {B.}~\bibnamefont
  {Holdom}},\ }\href {\doibase 10.1016/0370-2693(86)91377-8} {\bibfield
  {journal} {\bibinfo  {journal} {Phys. Lett. B}\ }\textbf {\bibinfo {volume}
  {166}},\ \bibinfo {pages} {196} (\bibinfo {year}
  {1986}{\natexlab{a}})}\BibitemShut {NoStop}%
\bibitem [{\citenamefont {Holdom}(1986{\natexlab{b}})}]{Holdom:1986eq}%
  \BibitemOpen
  \bibfield  {author} {\bibinfo {author} {\bibfnamefont {B.}~\bibnamefont
  {Holdom}},\ }\href {\doibase 10.1016/0370-2693(86)90470-3} {\bibfield
  {journal} {\bibinfo  {journal} {Phys. Lett. B}\ }\textbf {\bibinfo {volume}
  {178}},\ \bibinfo {pages} {65} (\bibinfo {year}
  {1986}{\natexlab{b}})}\BibitemShut {NoStop}%
\bibitem [{\citenamefont {Nelson}\ and\ \citenamefont
  {Scholtz}(2011)}]{Nelson:2011sf}%
  \BibitemOpen
  \bibfield  {author} {\bibinfo {author} {\bibfnamefont {A.~E.}\ \bibnamefont
  {Nelson}}\ and\ \bibinfo {author} {\bibfnamefont {J.}~\bibnamefont
  {Scholtz}},\ }\href {\doibase 10.1103/PhysRevD.84.103501} {\bibfield
  {journal} {\bibinfo  {journal} {Phys. Rev. D}\ }\textbf {\bibinfo {volume}
  {84}},\ \bibinfo {pages} {103501} (\bibinfo {year} {2011})},\ \Eprint
  {http://arxiv.org/abs/1105.2812} {arXiv:1105.2812 [hep-ph]} \BibitemShut
  {NoStop}%
\bibitem [{\citenamefont {Arias}\ \emph {et~al.}(2012)\citenamefont {Arias},
  \citenamefont {Cadamuro}, \citenamefont {Goodsell}, \citenamefont {Jaeckel},
  \citenamefont {Redondo},\ and\ \citenamefont {Ringwald}}]{Arias:2012az}%
  \BibitemOpen
  \bibfield  {author} {\bibinfo {author} {\bibfnamefont {P.}~\bibnamefont
  {Arias}}, \bibinfo {author} {\bibfnamefont {D.}~\bibnamefont {Cadamuro}},
  \bibinfo {author} {\bibfnamefont {M.}~\bibnamefont {Goodsell}}, \bibinfo
  {author} {\bibfnamefont {J.}~\bibnamefont {Jaeckel}}, \bibinfo {author}
  {\bibfnamefont {J.}~\bibnamefont {Redondo}}, \ and\ \bibinfo {author}
  {\bibfnamefont {A.}~\bibnamefont {Ringwald}},\ }\href {\doibase
  10.1088/1475-7516/2012/06/013} {\bibfield  {journal} {\bibinfo  {journal}
  {JCAP}\ }\textbf {\bibinfo {volume} {06}},\ \bibinfo {pages} {013} (\bibinfo
  {year} {2012})},\ \Eprint {http://arxiv.org/abs/1201.5902} {arXiv:1201.5902
  [hep-ph]} \BibitemShut {NoStop}%
\bibitem [{\citenamefont {Graham}\ \emph {et~al.}(2016)\citenamefont {Graham},
  \citenamefont {Mardon},\ and\ \citenamefont {Rajendran}}]{Graham:2015rva}%
  \BibitemOpen
  \bibfield  {author} {\bibinfo {author} {\bibfnamefont {P.~W.}\ \bibnamefont
  {Graham}}, \bibinfo {author} {\bibfnamefont {J.}~\bibnamefont {Mardon}}, \
  and\ \bibinfo {author} {\bibfnamefont {S.}~\bibnamefont {Rajendran}},\ }\href
  {\doibase 10.1103/PhysRevD.93.103520} {\bibfield  {journal} {\bibinfo
  {journal} {Phys. Rev. D}\ }\textbf {\bibinfo {volume} {93}},\ \bibinfo
  {pages} {103520} (\bibinfo {year} {2016})},\ \Eprint
  {http://arxiv.org/abs/1504.02102} {arXiv:1504.02102 [hep-ph]} \BibitemShut
  {NoStop}%
\bibitem [{\citenamefont {Fabbrichesi}\ \emph {et~al.}(2020)\citenamefont
  {Fabbrichesi}, \citenamefont {Gabrielli},\ and\ \citenamefont
  {Lanfranchi}}]{Fabbrichesi:2020wbt}%
  \BibitemOpen
  \bibfield  {author} {\bibinfo {author} {\bibfnamefont {M.}~\bibnamefont
  {Fabbrichesi}}, \bibinfo {author} {\bibfnamefont {E.}~\bibnamefont
  {Gabrielli}}, \ and\ \bibinfo {author} {\bibfnamefont {G.}~\bibnamefont
  {Lanfranchi}},\ }\href {\doibase 10.1007/978-3-030-62519-1} {\  (\bibinfo
  {year} {2020}),\ 10.1007/978-3-030-62519-1},\ \Eprint
  {http://arxiv.org/abs/2005.01515} {arXiv:2005.01515 [hep-ph]} \BibitemShut
  {NoStop}%
\bibitem [{\citenamefont {Arias}\ \emph {et~al.}(2015)\citenamefont {Arias},
  \citenamefont {Arza}, \citenamefont {D\"obrich}, \citenamefont {Gamboa},\
  and\ \citenamefont {M\'endez}}]{Arias:2014ela}%
  \BibitemOpen
  \bibfield  {author} {\bibinfo {author} {\bibfnamefont {P.}~\bibnamefont
  {Arias}}, \bibinfo {author} {\bibfnamefont {A.}~\bibnamefont {Arza}},
  \bibinfo {author} {\bibfnamefont {B.}~\bibnamefont {D\"obrich}}, \bibinfo
  {author} {\bibfnamefont {J.}~\bibnamefont {Gamboa}}, \ and\ \bibinfo {author}
  {\bibfnamefont {F.}~\bibnamefont {M\'endez}},\ }\href {\doibase
  10.1140/epjc/s10052-015-3536-0} {\bibfield  {journal} {\bibinfo  {journal}
  {Eur. Phys. J. C}\ }\textbf {\bibinfo {volume} {75}},\ \bibinfo {pages} {310}
  (\bibinfo {year} {2015})},\ \Eprint {http://arxiv.org/abs/1411.4986}
  {arXiv:1411.4986 [hep-ph]} \BibitemShut {NoStop}%
\bibitem [{\citenamefont {Chaudhuri}\ \emph {et~al.}(2015)\citenamefont
  {Chaudhuri}, \citenamefont {Graham}, \citenamefont {Irwin}, \citenamefont
  {Mardon}, \citenamefont {Rajendran},\ and\ \citenamefont
  {Zhao}}]{Chaudhuri:2014dla}%
  \BibitemOpen
  \bibfield  {author} {\bibinfo {author} {\bibfnamefont {S.}~\bibnamefont
  {Chaudhuri}}, \bibinfo {author} {\bibfnamefont {P.~W.}\ \bibnamefont
  {Graham}}, \bibinfo {author} {\bibfnamefont {K.}~\bibnamefont {Irwin}},
  \bibinfo {author} {\bibfnamefont {J.}~\bibnamefont {Mardon}}, \bibinfo
  {author} {\bibfnamefont {S.}~\bibnamefont {Rajendran}}, \ and\ \bibinfo
  {author} {\bibfnamefont {Y.}~\bibnamefont {Zhao}},\ }\href {\doibase
  10.1103/PhysRevD.92.075012} {\bibfield  {journal} {\bibinfo  {journal} {Phys.
  Rev. D}\ }\textbf {\bibinfo {volume} {92}},\ \bibinfo {pages} {075012}
  (\bibinfo {year} {2015})},\ \Eprint {http://arxiv.org/abs/1411.7382}
  {arXiv:1411.7382 [hep-ph]} \BibitemShut {NoStop}%
\bibitem [{\citenamefont {Nguyen}\ \emph {et~al.}(2019)\citenamefont {Nguyen},
  \citenamefont {Lobanov},\ and\ \citenamefont {Horns}}]{Nguyen:2019xuh}%
  \BibitemOpen
  \bibfield  {author} {\bibinfo {author} {\bibfnamefont {L.~H.}\ \bibnamefont
  {Nguyen}}, \bibinfo {author} {\bibfnamefont {A.}~\bibnamefont {Lobanov}}, \
  and\ \bibinfo {author} {\bibfnamefont {D.}~\bibnamefont {Horns}},\ }\href
  {\doibase 10.1088/1475-7516/2019/10/014} {\bibfield  {journal} {\bibinfo
  {journal} {JCAP}\ }\textbf {\bibinfo {volume} {10}},\ \bibinfo {pages} {014}
  (\bibinfo {year} {2019})},\ \Eprint {http://arxiv.org/abs/1907.12449}
  {arXiv:1907.12449 [hep-ex]} \BibitemShut {NoStop}%
\bibitem [{\citenamefont {Cervantes}\ \emph
  {et~al.}(2022{\natexlab{a}})\citenamefont {Cervantes} \emph
  {et~al.}}]{Cervantes:2022epl}%
  \BibitemOpen
  \bibfield  {author} {\bibinfo {author} {\bibfnamefont {R.}~\bibnamefont
  {Cervantes}} \emph {et~al.},\ }\href {\doibase 10.1103/PhysRevD.106.102002}
  {\bibfield  {journal} {\bibinfo  {journal} {Phys. Rev. D}\ }\textbf {\bibinfo
  {volume} {106}},\ \bibinfo {pages} {102002} (\bibinfo {year}
  {2022}{\natexlab{a}})},\ \Eprint {http://arxiv.org/abs/2204.09475}
  {arXiv:2204.09475 [hep-ex]} \BibitemShut {NoStop}%
\bibitem [{\citenamefont {Cervantes}\ \emph
  {et~al.}(2022{\natexlab{b}})\citenamefont {Cervantes}, \citenamefont
  {Braggio}, \citenamefont {Giaccone}, \citenamefont {Frolov}, \citenamefont
  {Grassellino}, \citenamefont {Harnik}, \citenamefont {Melnychuk},
  \citenamefont {Pilipenko}, \citenamefont {Posen},\ and\ \citenamefont
  {Romanenko}}]{Cervantes:2022gtv}%
  \BibitemOpen
  \bibfield  {author} {\bibinfo {author} {\bibfnamefont {R.}~\bibnamefont
  {Cervantes}}, \bibinfo {author} {\bibfnamefont {C.}~\bibnamefont {Braggio}},
  \bibinfo {author} {\bibfnamefont {B.}~\bibnamefont {Giaccone}}, \bibinfo
  {author} {\bibfnamefont {D.}~\bibnamefont {Frolov}}, \bibinfo {author}
  {\bibfnamefont {A.}~\bibnamefont {Grassellino}}, \bibinfo {author}
  {\bibfnamefont {R.}~\bibnamefont {Harnik}}, \bibinfo {author} {\bibfnamefont
  {O.}~\bibnamefont {Melnychuk}}, \bibinfo {author} {\bibfnamefont
  {R.}~\bibnamefont {Pilipenko}}, \bibinfo {author} {\bibfnamefont
  {S.}~\bibnamefont {Posen}}, \ and\ \bibinfo {author} {\bibfnamefont
  {A.}~\bibnamefont {Romanenko}},\ }\href@noop {} {\  (\bibinfo {year}
  {2022}{\natexlab{b}})},\ \Eprint {http://arxiv.org/abs/2208.03183}
  {arXiv:2208.03183 [hep-ex]} \BibitemShut {NoStop}%
\bibitem [{\citenamefont {McAllister}\ \emph {et~al.}(2024)\citenamefont
  {McAllister}, \citenamefont {Quiskamp}, \citenamefont {O'Hare}, \citenamefont
  {Altin}, \citenamefont {Ivanov}, \citenamefont {Goryachev},\ and\
  \citenamefont {Tobar}}]{McAllister:2022ibe}%
  \BibitemOpen
  \bibfield  {author} {\bibinfo {author} {\bibfnamefont {B.~T.}\ \bibnamefont
  {McAllister}}, \bibinfo {author} {\bibfnamefont {A.}~\bibnamefont
  {Quiskamp}}, \bibinfo {author} {\bibfnamefont {C.~A.~J.}\ \bibnamefont
  {O'Hare}}, \bibinfo {author} {\bibfnamefont {P.}~\bibnamefont {Altin}},
  \bibinfo {author} {\bibfnamefont {E.~N.}\ \bibnamefont {Ivanov}}, \bibinfo
  {author} {\bibfnamefont {M.}~\bibnamefont {Goryachev}}, \ and\ \bibinfo
  {author} {\bibfnamefont {M.~E.}\ \bibnamefont {Tobar}},\ }\href {\doibase
  10.1002/andp.202200622} {\bibfield  {journal} {\bibinfo  {journal} {Annalen
  Phys.}\ }\textbf {\bibinfo {volume} {536}},\ \bibinfo {pages} {2200622}
  (\bibinfo {year} {2024})},\ \Eprint {http://arxiv.org/abs/2212.01971}
  {arXiv:2212.01971 [hep-ph]} \BibitemShut {NoStop}%
\bibitem [{\citenamefont {Cervantes}\ \emph
  {et~al.}(2022{\natexlab{c}})\citenamefont {Cervantes} \emph
  {et~al.}}]{Cervantes:2022yzp}%
  \BibitemOpen
  \bibfield  {author} {\bibinfo {author} {\bibfnamefont {R.}~\bibnamefont
  {Cervantes}} \emph {et~al.},\ }\href {\doibase
  10.1103/PhysRevLett.129.201301} {\bibfield  {journal} {\bibinfo  {journal}
  {Phys. Rev. Lett.}\ }\textbf {\bibinfo {volume} {129}},\ \bibinfo {pages}
  {201301} (\bibinfo {year} {2022}{\natexlab{c}})},\ \Eprint
  {http://arxiv.org/abs/2204.03818} {arXiv:2204.03818 [hep-ex]} \BibitemShut
  {NoStop}%
\bibitem [{\citenamefont {Ramanathan}\ \emph {et~al.}(2023)\citenamefont
  {Ramanathan}, \citenamefont {Klimovich}, \citenamefont {Basu~Thakur},
  \citenamefont {Eom}, \citenamefont {LeDuc}, \citenamefont {Shu},
  \citenamefont {Beyer},\ and\ \citenamefont {Day}}]{Ramanathan:2022egk}%
  \BibitemOpen
  \bibfield  {author} {\bibinfo {author} {\bibfnamefont {K.}~\bibnamefont
  {Ramanathan}}, \bibinfo {author} {\bibfnamefont {N.}~\bibnamefont
  {Klimovich}}, \bibinfo {author} {\bibfnamefont {R.}~\bibnamefont
  {Basu~Thakur}}, \bibinfo {author} {\bibfnamefont {B.~H.}\ \bibnamefont
  {Eom}}, \bibinfo {author} {\bibfnamefont {H.~G.}\ \bibnamefont {LeDuc}},
  \bibinfo {author} {\bibfnamefont {S.}~\bibnamefont {Shu}}, \bibinfo {author}
  {\bibfnamefont {A.~D.}\ \bibnamefont {Beyer}}, \ and\ \bibinfo {author}
  {\bibfnamefont {P.~K.}\ \bibnamefont {Day}},\ }\href {\doibase
  10.1103/PhysRevLett.130.231001} {\bibfield  {journal} {\bibinfo  {journal}
  {Phys. Rev. Lett.}\ }\textbf {\bibinfo {volume} {130}},\ \bibinfo {pages}
  {231001} (\bibinfo {year} {2023})},\ \Eprint
  {http://arxiv.org/abs/2209.03419} {arXiv:2209.03419 [astro-ph.CO]}
  \BibitemShut {NoStop}%
\bibitem [{\citenamefont {Schneemann}\ \emph {et~al.}(2023)\citenamefont
  {Schneemann}, \citenamefont {Schmieden},\ and\ \citenamefont
  {Schott}}]{Schneemann:2023bqc}%
  \BibitemOpen
  \bibfield  {author} {\bibinfo {author} {\bibfnamefont {T.}~\bibnamefont
  {Schneemann}}, \bibinfo {author} {\bibfnamefont {K.}~\bibnamefont
  {Schmieden}}, \ and\ \bibinfo {author} {\bibfnamefont {M.}~\bibnamefont
  {Schott}},\ }\href@noop {} {\  (\bibinfo {year} {2023})},\ \Eprint
  {http://arxiv.org/abs/2308.08337} {arXiv:2308.08337 [hep-ex]} \BibitemShut
  {NoStop}%
\bibitem [{\citenamefont {Tang}\ \emph {et~al.}(2024)\citenamefont {Tang} \emph
  {et~al.}}]{Tang:2023oid}%
  \BibitemOpen
  \bibfield  {author} {\bibinfo {author} {\bibfnamefont {Z.}~\bibnamefont
  {Tang}} \emph {et~al.} (\bibinfo {collaboration} {SHANHE}),\ }\href {\doibase
  10.1103/PhysRevLett.133.021005} {\bibfield  {journal} {\bibinfo  {journal}
  {Phys. Rev. Lett.}\ }\textbf {\bibinfo {volume} {133}},\ \bibinfo {pages}
  {021005} (\bibinfo {year} {2024})},\ \Eprint
  {http://arxiv.org/abs/2305.09711} {arXiv:2305.09711 [hep-ex]} \BibitemShut
  {NoStop}%
\bibitem [{\citenamefont {He}\ \emph {et~al.}(2024)\citenamefont {He} \emph
  {et~al.}}]{He:2024ytp}%
  \BibitemOpen
  \bibfield  {author} {\bibinfo {author} {\bibfnamefont {D.}~\bibnamefont {He}}
  \emph {et~al.} (\bibinfo {collaboration} {APEX}),\ }\href {\doibase
  10.1103/PhysRevD.110.L021101} {\bibfield  {journal} {\bibinfo  {journal}
  {Phys. Rev. D}\ }\textbf {\bibinfo {volume} {110}},\ \bibinfo {pages}
  {L021101} (\bibinfo {year} {2024})},\ \Eprint
  {http://arxiv.org/abs/2404.00908} {arXiv:2404.00908 [hep-ex]} \BibitemShut
  {NoStop}%
\bibitem [{\citenamefont {Dirac}(1931)}]{Dirac:1931kp}%
  \BibitemOpen
  \bibfield  {author} {\bibinfo {author} {\bibfnamefont {P.~A.~M.}\
  \bibnamefont {Dirac}},\ }\href {\doibase 10.1098/rspa.1931.0130} {\bibfield
  {journal} {\bibinfo  {journal} {Proc. Roy. Soc. Lond. A}\ }\textbf {\bibinfo
  {volume} {133}},\ \bibinfo {pages} {60} (\bibinfo {year} {1931})}\BibitemShut
  {NoStop}%
\bibitem [{\citenamefont {Wu}\ and\ \citenamefont {Yang}(1975)}]{Wu:1975es}%
  \BibitemOpen
  \bibfield  {author} {\bibinfo {author} {\bibfnamefont {T.~T.}\ \bibnamefont
  {Wu}}\ and\ \bibinfo {author} {\bibfnamefont {C.~N.}\ \bibnamefont {Yang}},\
  }\href {\doibase 10.1103/PhysRevD.12.3845} {\bibfield  {journal} {\bibinfo
  {journal} {Phys. Rev. D}\ }\textbf {\bibinfo {volume} {12}},\ \bibinfo
  {pages} {3845} (\bibinfo {year} {1975})}\BibitemShut {NoStop}%
\bibitem [{\citenamefont {'t~Hooft}(1974)}]{tHooft:1974kcl}%
  \BibitemOpen
  \bibfield  {author} {\bibinfo {author} {\bibfnamefont {G.}~\bibnamefont
  {'t~Hooft}},\ }\href {\doibase 10.1016/0550-3213(74)90486-6} {\bibfield
  {journal} {\bibinfo  {journal} {Nucl. Phys. B}\ }\textbf {\bibinfo {volume}
  {79}},\ \bibinfo {pages} {276} (\bibinfo {year} {1974})}\BibitemShut
  {NoStop}%
\bibitem [{\citenamefont {Polyakov}(1974)}]{Polyakov:1974ek}%
  \BibitemOpen
  \bibfield  {author} {\bibinfo {author} {\bibfnamefont {A.~M.}\ \bibnamefont
  {Polyakov}},\ }\href@noop {} {\bibfield  {journal} {\bibinfo  {journal} {JETP
  Lett.}\ }\textbf {\bibinfo {volume} {20}},\ \bibinfo {pages} {194} (\bibinfo
  {year} {1974})}\BibitemShut {NoStop}%
\bibitem [{\citenamefont {Cho}\ and\ \citenamefont
  {Maison}(1997)}]{Cho:1996qd}%
  \BibitemOpen
  \bibfield  {author} {\bibinfo {author} {\bibfnamefont {Y.~M.}\ \bibnamefont
  {Cho}}\ and\ \bibinfo {author} {\bibfnamefont {D.}~\bibnamefont {Maison}},\
  }\href {\doibase 10.1016/S0370-2693(96)01492-X} {\bibfield  {journal}
  {\bibinfo  {journal} {Phys. Lett. B}\ }\textbf {\bibinfo {volume} {391}},\
  \bibinfo {pages} {360} (\bibinfo {year} {1997})},\ \Eprint
  {http://arxiv.org/abs/hep-th/9601028} {arXiv:hep-th/9601028} \BibitemShut
  {NoStop}%
\bibitem [{\citenamefont {Hung}(2021)}]{Hung:2020vuo}%
  \BibitemOpen
  \bibfield  {author} {\bibinfo {author} {\bibfnamefont {P.~Q.}\ \bibnamefont
  {Hung}},\ }\href {\doibase 10.1016/j.nuclphysb.2020.115278} {\bibfield
  {journal} {\bibinfo  {journal} {Nucl. Phys. B}\ }\textbf {\bibinfo {volume}
  {962}},\ \bibinfo {pages} {115278} (\bibinfo {year} {2021})},\ \Eprint
  {http://arxiv.org/abs/2003.02794} {arXiv:2003.02794 [hep-ph]} \BibitemShut
  {NoStop}%
\bibitem [{\citenamefont {Alexandre}\ and\ \citenamefont
  {Mavromatos}(2019)}]{Alexandre:2019iub}%
  \BibitemOpen
  \bibfield  {author} {\bibinfo {author} {\bibfnamefont {J.}~\bibnamefont
  {Alexandre}}\ and\ \bibinfo {author} {\bibfnamefont {N.~E.}\ \bibnamefont
  {Mavromatos}},\ }\href {\doibase 10.1103/PhysRevD.100.096005} {\bibfield
  {journal} {\bibinfo  {journal} {Phys. Rev. D}\ }\textbf {\bibinfo {volume}
  {100}},\ \bibinfo {pages} {096005} (\bibinfo {year} {2019})},\ \Eprint
  {http://arxiv.org/abs/1906.08738} {arXiv:1906.08738 [hep-ph]} \BibitemShut
  {NoStop}%
\bibitem [{\citenamefont {Ellis}\ \emph {et~al.}(2016)\citenamefont {Ellis},
  \citenamefont {Mavromatos},\ and\ \citenamefont {You}}]{Ellis:2016glu}%
  \BibitemOpen
  \bibfield  {author} {\bibinfo {author} {\bibfnamefont {J.}~\bibnamefont
  {Ellis}}, \bibinfo {author} {\bibfnamefont {N.~E.}\ \bibnamefont
  {Mavromatos}}, \ and\ \bibinfo {author} {\bibfnamefont {T.}~\bibnamefont
  {You}},\ }\href {\doibase 10.1016/j.physletb.2016.02.048} {\bibfield
  {journal} {\bibinfo  {journal} {Phys. Lett. B}\ }\textbf {\bibinfo {volume}
  {756}},\ \bibinfo {pages} {29} (\bibinfo {year} {2016})},\ \Eprint
  {http://arxiv.org/abs/1602.01745} {arXiv:1602.01745 [hep-ph]} \BibitemShut
  {NoStop}%
\bibitem [{\citenamefont {Lazarides}\ and\ \citenamefont
  {Shafi}(2021)}]{Lazarides:2021los}%
  \BibitemOpen
  \bibfield  {author} {\bibinfo {author} {\bibfnamefont {G.}~\bibnamefont
  {Lazarides}}\ and\ \bibinfo {author} {\bibfnamefont {Q.}~\bibnamefont
  {Shafi}},\ }\href {\doibase 10.1103/PhysRevD.103.095021} {\bibfield
  {journal} {\bibinfo  {journal} {Phys. Rev. D}\ }\textbf {\bibinfo {volume}
  {103}},\ \bibinfo {pages} {095021} (\bibinfo {year} {2021})},\ \Eprint
  {http://arxiv.org/abs/2102.07124} {arXiv:2102.07124 [hep-ph]} \BibitemShut
  {NoStop}%
\bibitem [{\citenamefont {Schwinger}(1966{\natexlab{a}})}]{Schwinger:1966nj}%
  \BibitemOpen
  \bibfield  {author} {\bibinfo {author} {\bibfnamefont {J.~S.}\ \bibnamefont
  {Schwinger}},\ }\href {\doibase 10.1103/PhysRev.144.1087} {\bibfield
  {journal} {\bibinfo  {journal} {Phys. Rev.}\ }\textbf {\bibinfo {volume}
  {144}},\ \bibinfo {pages} {1087} (\bibinfo {year}
  {1966}{\natexlab{a}})}\BibitemShut {NoStop}%
\bibitem [{\citenamefont {Zwanziger}(1968)}]{Zwanziger:1968rs}%
  \BibitemOpen
  \bibfield  {author} {\bibinfo {author} {\bibfnamefont {D.}~\bibnamefont
  {Zwanziger}},\ }\href {\doibase 10.1103/PhysRev.176.1489} {\bibfield
  {journal} {\bibinfo  {journal} {Phys. Rev.}\ }\textbf {\bibinfo {volume}
  {176}},\ \bibinfo {pages} {1489} (\bibinfo {year} {1968})}\BibitemShut
  {NoStop}%
\bibitem [{\citenamefont {Zwanziger}(1971)}]{Zwanziger:1970hk}%
  \BibitemOpen
  \bibfield  {author} {\bibinfo {author} {\bibfnamefont {D.}~\bibnamefont
  {Zwanziger}},\ }\href {\doibase 10.1103/PhysRevD.3.880} {\bibfield  {journal}
  {\bibinfo  {journal} {Phys. Rev. D}\ }\textbf {\bibinfo {volume} {3}},\
  \bibinfo {pages} {880} (\bibinfo {year} {1971})}\BibitemShut {NoStop}%
\bibitem [{\citenamefont {Brandt}\ \emph {et~al.}(1978)\citenamefont {Brandt},
  \citenamefont {Neri},\ and\ \citenamefont {Zwanziger}}]{Brandt:1977be}%
  \BibitemOpen
  \bibfield  {author} {\bibinfo {author} {\bibfnamefont {R.~A.}\ \bibnamefont
  {Brandt}}, \bibinfo {author} {\bibfnamefont {F.}~\bibnamefont {Neri}}, \ and\
  \bibinfo {author} {\bibfnamefont {D.}~\bibnamefont {Zwanziger}},\ }\href
  {\doibase 10.1103/PhysRevLett.40.147} {\bibfield  {journal} {\bibinfo
  {journal} {Phys. Rev. Lett.}\ }\textbf {\bibinfo {volume} {40}},\ \bibinfo
  {pages} {147} (\bibinfo {year} {1978})}\BibitemShut {NoStop}%
\bibitem [{\citenamefont {Brandt}\ \emph {et~al.}(1979)\citenamefont {Brandt},
  \citenamefont {Neri},\ and\ \citenamefont {Zwanziger}}]{Brandt:1978wc}%
  \BibitemOpen
  \bibfield  {author} {\bibinfo {author} {\bibfnamefont {R.~A.}\ \bibnamefont
  {Brandt}}, \bibinfo {author} {\bibfnamefont {F.}~\bibnamefont {Neri}}, \ and\
  \bibinfo {author} {\bibfnamefont {D.}~\bibnamefont {Zwanziger}},\ }\href
  {\doibase 10.1103/PhysRevD.19.1153} {\bibfield  {journal} {\bibinfo
  {journal} {Phys. Rev. D}\ }\textbf {\bibinfo {volume} {19}},\ \bibinfo
  {pages} {1153} (\bibinfo {year} {1979})}\BibitemShut {NoStop}%
\bibitem [{\citenamefont {Sokolov}\ and\ \citenamefont
  {Ringwald}(2023)}]{Sokolov:2023pos}%
  \BibitemOpen
  \bibfield  {author} {\bibinfo {author} {\bibfnamefont {A.~V.}\ \bibnamefont
  {Sokolov}}\ and\ \bibinfo {author} {\bibfnamefont {A.}~\bibnamefont
  {Ringwald}},\ }\href {\doibase 10.1002/andp.202300112} {\bibfield  {journal}
  {\bibinfo  {journal} {Annalen Phys.}\ }\textbf {\bibinfo {volume} {2023}}
  (\bibinfo {year} {2023}),\ 10.1002/andp.202300112},\ \Eprint
  {http://arxiv.org/abs/2303.10170} {arXiv:2303.10170 [hep-ph]} \BibitemShut
  {NoStop}%
\bibitem [{\citenamefont {Sokolov}\ and\ \citenamefont
  {Ringwald}(2022)}]{Sokolov:2022fvs}%
  \BibitemOpen
  \bibfield  {author} {\bibinfo {author} {\bibfnamefont {A.~V.}\ \bibnamefont
  {Sokolov}}\ and\ \bibinfo {author} {\bibfnamefont {A.}~\bibnamefont
  {Ringwald}},\ }\href@noop {} {\  (\bibinfo {year} {2022})},\ \Eprint
  {http://arxiv.org/abs/2205.02605} {arXiv:2205.02605 [hep-ph]} \BibitemShut
  {NoStop}%
\bibitem [{\citenamefont {Heidenreich}\ \emph {et~al.}(2024)\citenamefont
  {Heidenreich}, \citenamefont {McNamara},\ and\ \citenamefont
  {Reece}}]{Heidenreich:2023pbi}%
  \BibitemOpen
  \bibfield  {author} {\bibinfo {author} {\bibfnamefont {B.}~\bibnamefont
  {Heidenreich}}, \bibinfo {author} {\bibfnamefont {J.}~\bibnamefont
  {McNamara}}, \ and\ \bibinfo {author} {\bibfnamefont {M.}~\bibnamefont
  {Reece}},\ }\href {\doibase 10.1007/JHEP01(2024)120} {\bibfield  {journal}
  {\bibinfo  {journal} {JHEP}\ }\textbf {\bibinfo {volume} {01}},\ \bibinfo
  {pages} {120} (\bibinfo {year} {2024})},\ \Eprint
  {http://arxiv.org/abs/2309.07951} {arXiv:2309.07951 [hep-ph]} \BibitemShut
  {NoStop}%
\bibitem [{\citenamefont {Li}\ and\ \citenamefont
  {Zhang}(2023{\natexlab{a}})}]{Li:2023zcp}%
  \BibitemOpen
  \bibfield  {author} {\bibinfo {author} {\bibfnamefont {T.}~\bibnamefont
  {Li}}\ and\ \bibinfo {author} {\bibfnamefont {R.-J.}\ \bibnamefont {Zhang}},\
  }\href@noop {} {\  (\bibinfo {year} {2023}{\natexlab{a}})},\ \Eprint
  {http://arxiv.org/abs/2312.01355} {arXiv:2312.01355 [hep-ph]} \BibitemShut
  {NoStop}%
\bibitem [{\citenamefont {Li}\ \emph {et~al.}(2023)\citenamefont {Li},
  \citenamefont {Zhang},\ and\ \citenamefont {Dai}}]{Li:2022oel}%
  \BibitemOpen
  \bibfield  {author} {\bibinfo {author} {\bibfnamefont {T.}~\bibnamefont
  {Li}}, \bibinfo {author} {\bibfnamefont {R.-J.}\ \bibnamefont {Zhang}}, \
  and\ \bibinfo {author} {\bibfnamefont {C.-J.}\ \bibnamefont {Dai}},\ }\href
  {\doibase 10.1007/JHEP03(2023)088} {\bibfield  {journal} {\bibinfo  {journal}
  {JHEP}\ }\textbf {\bibinfo {volume} {03}},\ \bibinfo {pages} {088} (\bibinfo
  {year} {2023})},\ \Eprint {http://arxiv.org/abs/2211.06847} {arXiv:2211.06847
  [hep-ph]} \BibitemShut {NoStop}%
\bibitem [{\citenamefont {Tobar}\ \emph {et~al.}(2024)\citenamefont {Tobar},
  \citenamefont {Thomson}, \citenamefont {McAllister}, \citenamefont
  {Goryachev}, \citenamefont {Sokolov},\ and\ \citenamefont
  {Ringwald}}]{Tobar:2022rko}%
  \BibitemOpen
  \bibfield  {author} {\bibinfo {author} {\bibfnamefont {M.~E.}\ \bibnamefont
  {Tobar}}, \bibinfo {author} {\bibfnamefont {C.~A.}\ \bibnamefont {Thomson}},
  \bibinfo {author} {\bibfnamefont {B.~T.}\ \bibnamefont {McAllister}},
  \bibinfo {author} {\bibfnamefont {M.}~\bibnamefont {Goryachev}}, \bibinfo
  {author} {\bibfnamefont {A.~V.}\ \bibnamefont {Sokolov}}, \ and\ \bibinfo
  {author} {\bibfnamefont {A.}~\bibnamefont {Ringwald}},\ }\href {\doibase
  10.1002/andp.202200594} {\bibfield  {journal} {\bibinfo  {journal} {Annalen
  Phys.}\ }\textbf {\bibinfo {volume} {536}},\ \bibinfo {pages} {2200594}
  (\bibinfo {year} {2024})},\ \Eprint {http://arxiv.org/abs/2211.09637}
  {arXiv:2211.09637 [hep-ph]} \BibitemShut {NoStop}%
\bibitem [{\citenamefont {Li}\ \emph {et~al.}(2024)\citenamefont {Li},
  \citenamefont {Dai},\ and\ \citenamefont {Zhang}}]{Li:2023kfh}%
  \BibitemOpen
  \bibfield  {author} {\bibinfo {author} {\bibfnamefont {T.}~\bibnamefont
  {Li}}, \bibinfo {author} {\bibfnamefont {C.-J.}\ \bibnamefont {Dai}}, \ and\
  \bibinfo {author} {\bibfnamefont {R.-J.}\ \bibnamefont {Zhang}},\ }\href
  {\doibase 10.1103/PhysRevD.109.015026} {\bibfield  {journal} {\bibinfo
  {journal} {Phys. Rev. D}\ }\textbf {\bibinfo {volume} {109}},\ \bibinfo
  {pages} {015026} (\bibinfo {year} {2024})},\ \Eprint
  {http://arxiv.org/abs/2304.12525} {arXiv:2304.12525 [hep-ph]} \BibitemShut
  {NoStop}%
\bibitem [{\citenamefont {Li}\ and\ \citenamefont
  {Zhang}(2023{\natexlab{b}})}]{Li:2023aow}%
  \BibitemOpen
  \bibfield  {author} {\bibinfo {author} {\bibfnamefont {T.}~\bibnamefont
  {Li}}\ and\ \bibinfo {author} {\bibfnamefont {R.-J.}\ \bibnamefont {Zhang}},\
  }\href {\doibase 10.1088/1674-1137/ad0620} {\bibfield  {journal} {\bibinfo
  {journal} {Chin. Phys. C}\ }\textbf {\bibinfo {volume} {47}},\ \bibinfo
  {pages} {123104} (\bibinfo {year} {2023}{\natexlab{b}})},\ \Eprint
  {http://arxiv.org/abs/2305.01344} {arXiv:2305.01344 [hep-ph]} \BibitemShut
  {NoStop}%
\bibitem [{\citenamefont {Tobar}\ \emph {et~al.}(2023)\citenamefont {Tobar},
  \citenamefont {Sokolov}, \citenamefont {Ringwald},\ and\ \citenamefont
  {Goryachev}}]{Tobar:2023rga}%
  \BibitemOpen
  \bibfield  {author} {\bibinfo {author} {\bibfnamefont {M.~E.}\ \bibnamefont
  {Tobar}}, \bibinfo {author} {\bibfnamefont {A.~V.}\ \bibnamefont {Sokolov}},
  \bibinfo {author} {\bibfnamefont {A.}~\bibnamefont {Ringwald}}, \ and\
  \bibinfo {author} {\bibfnamefont {M.}~\bibnamefont {Goryachev}},\ }\href
  {\doibase 10.1103/PhysRevD.108.035024} {\bibfield  {journal} {\bibinfo
  {journal} {Phys. Rev. D}\ }\textbf {\bibinfo {volume} {108}},\ \bibinfo
  {pages} {035024} (\bibinfo {year} {2023})},\ \Eprint
  {http://arxiv.org/abs/2306.13320} {arXiv:2306.13320 [hep-ph]} \BibitemShut
  {NoStop}%
\bibitem [{\citenamefont {Patkos}(2023)}]{Patkos:2023lof}%
  \BibitemOpen
  \bibfield  {author} {\bibinfo {author} {\bibfnamefont {A.}~\bibnamefont
  {Patkos}},\ }\href {\doibase 10.1142/S0217732323501377} {\bibfield  {journal}
  {\bibinfo  {journal} {Mod. Phys. Lett. A}\ }\textbf {\bibinfo {volume}
  {38}},\ \bibinfo {pages} {2350137} (\bibinfo {year} {2023})},\ \Eprint
  {http://arxiv.org/abs/2309.05523} {arXiv:2309.05523 [hep-ph]} \BibitemShut
  {NoStop}%
\bibitem [{\citenamefont {Dai}\ \emph {et~al.}(2024)\citenamefont {Dai},
  \citenamefont {Li},\ and\ \citenamefont {Zhang}}]{Dai:2024dkr}%
  \BibitemOpen
  \bibfield  {author} {\bibinfo {author} {\bibfnamefont {C.-J.}\ \bibnamefont
  {Dai}}, \bibinfo {author} {\bibfnamefont {T.}~\bibnamefont {Li}}, \ and\
  \bibinfo {author} {\bibfnamefont {R.-J.}\ \bibnamefont {Zhang}},\ }\href@noop
  {} {\  (\bibinfo {year} {2024})},\ \Eprint {http://arxiv.org/abs/2401.14195}
  {arXiv:2401.14195 [hep-ph]} \BibitemShut {NoStop}%
\bibitem [{\citenamefont {Hook}\ and\ \citenamefont
  {Huang}(2017)}]{Hook:2017vyc}%
  \BibitemOpen
  \bibfield  {author} {\bibinfo {author} {\bibfnamefont {A.}~\bibnamefont
  {Hook}}\ and\ \bibinfo {author} {\bibfnamefont {J.}~\bibnamefont {Huang}},\
  }\href {\doibase 10.1103/PhysRevD.96.055010} {\bibfield  {journal} {\bibinfo
  {journal} {Phys. Rev. D}\ }\textbf {\bibinfo {volume} {96}},\ \bibinfo
  {pages} {055010} (\bibinfo {year} {2017})},\ \Eprint
  {http://arxiv.org/abs/1705.01107} {arXiv:1705.01107 [hep-ph]} \BibitemShut
  {NoStop}%
\bibitem [{\citenamefont {Terning}\ and\ \citenamefont
  {Verhaaren}(2018)}]{Terning:2018lsv}%
  \BibitemOpen
  \bibfield  {author} {\bibinfo {author} {\bibfnamefont {J.}~\bibnamefont
  {Terning}}\ and\ \bibinfo {author} {\bibfnamefont {C.~B.}\ \bibnamefont
  {Verhaaren}},\ }\href {\doibase 10.1007/JHEP12(2018)123} {\bibfield
  {journal} {\bibinfo  {journal} {JHEP}\ }\textbf {\bibinfo {volume} {12}},\
  \bibinfo {pages} {123} (\bibinfo {year} {2018})},\ \Eprint
  {http://arxiv.org/abs/1808.09459} {arXiv:1808.09459 [hep-th]} \BibitemShut
  {NoStop}%
\bibitem [{\citenamefont {Terning}\ and\ \citenamefont
  {Verhaaren}(2019)}]{Terning:2019bhg}%
  \BibitemOpen
  \bibfield  {author} {\bibinfo {author} {\bibfnamefont {J.}~\bibnamefont
  {Terning}}\ and\ \bibinfo {author} {\bibfnamefont {C.~B.}\ \bibnamefont
  {Verhaaren}},\ }\href {\doibase 10.1007/JHEP12(2019)152} {\bibfield
  {journal} {\bibinfo  {journal} {JHEP}\ }\textbf {\bibinfo {volume} {12}},\
  \bibinfo {pages} {152} (\bibinfo {year} {2019})},\ \Eprint
  {http://arxiv.org/abs/1906.00014} {arXiv:1906.00014 [hep-ph]} \BibitemShut
  {NoStop}%
\bibitem [{\citenamefont {Terning}\ and\ \citenamefont
  {Verhaaren}(2020)}]{Terning:2020dzg}%
  \BibitemOpen
  \bibfield  {author} {\bibinfo {author} {\bibfnamefont {J.}~\bibnamefont
  {Terning}}\ and\ \bibinfo {author} {\bibfnamefont {C.~B.}\ \bibnamefont
  {Verhaaren}},\ }\href {\doibase 10.1007/JHEP12(2020)153} {\bibfield
  {journal} {\bibinfo  {journal} {JHEP}\ }\textbf {\bibinfo {volume} {12}},\
  \bibinfo {pages} {153} (\bibinfo {year} {2020})},\ \Eprint
  {http://arxiv.org/abs/2010.02232} {arXiv:2010.02232 [hep-th]} \BibitemShut
  {NoStop}%
\bibitem [{\citenamefont {Schwinger}(1966{\natexlab{b}})}]{Schwinger:1966zz}%
  \BibitemOpen
  \bibfield  {author} {\bibinfo {author} {\bibfnamefont {J.}~\bibnamefont
  {Schwinger}},\ }\href {\doibase 10.1103/PhysRev.152.1219} {\bibfield
  {journal} {\bibinfo  {journal} {Phys. Rev.}\ }\textbf {\bibinfo {volume}
  {152}},\ \bibinfo {pages} {1219} (\bibinfo {year}
  {1966}{\natexlab{b}})}\BibitemShut {NoStop}%
\bibitem [{\citenamefont {Schwinger}(1967)}]{Schwinger:1967rg}%
  \BibitemOpen
  \bibfield  {author} {\bibinfo {author} {\bibfnamefont {J.~S.}\ \bibnamefont
  {Schwinger}},\ }\href {\doibase 10.1103/PhysRev.158.1391} {\bibfield
  {journal} {\bibinfo  {journal} {Phys. Rev.}\ }\textbf {\bibinfo {volume}
  {158}},\ \bibinfo {pages} {1391} (\bibinfo {year} {1967})}\BibitemShut
  {NoStop}%
\bibitem [{\citenamefont {Schwinger}(1968)}]{Schwinger:1968rq}%
  \BibitemOpen
  \bibfield  {author} {\bibinfo {author} {\bibfnamefont {J.~S.}\ \bibnamefont
  {Schwinger}},\ }\href {\doibase 10.1103/PhysRev.173.1536} {\bibfield
  {journal} {\bibinfo  {journal} {Phys. Rev.}\ }\textbf {\bibinfo {volume}
  {173}},\ \bibinfo {pages} {1536} (\bibinfo {year} {1968})}\BibitemShut
  {NoStop}%
\bibitem [{\citenamefont {Yan}(1966)}]{PhysRev.150.1349}%
  \BibitemOpen
  \bibfield  {author} {\bibinfo {author} {\bibfnamefont {T.-m.}\ \bibnamefont
  {Yan}},\ }\href {\doibase 10.1103/PhysRev.150.1349} {\bibfield  {journal}
  {\bibinfo  {journal} {Phys. Rev.}\ }\textbf {\bibinfo {volume} {150}},\
  \bibinfo {pages} {1349} (\bibinfo {year} {1966})}\BibitemShut {NoStop}%
\bibitem [{\citenamefont {Turner}(1990)}]{Turner:1990qx}%
  \BibitemOpen
  \bibfield  {author} {\bibinfo {author} {\bibfnamefont {M.~S.}\ \bibnamefont
  {Turner}},\ }\href {\doibase 10.1103/PhysRevD.42.3572} {\bibfield  {journal}
  {\bibinfo  {journal} {Phys. Rev. D}\ }\textbf {\bibinfo {volume} {42}},\
  \bibinfo {pages} {3572} (\bibinfo {year} {1990})}\BibitemShut {NoStop}%
\bibitem [{\citenamefont {Bartram}\ \emph {et~al.}(2021)\citenamefont {Bartram}
  \emph {et~al.}}]{ADMX:2021nhd}%
  \BibitemOpen
  \bibfield  {author} {\bibinfo {author} {\bibfnamefont {C.}~\bibnamefont
  {Bartram}} \emph {et~al.} (\bibinfo {collaboration} {ADMX}),\ }\href
  {\doibase 10.1103/PhysRevLett.127.261803} {\bibfield  {journal} {\bibinfo
  {journal} {Phys. Rev. Lett.}\ }\textbf {\bibinfo {volume} {127}},\ \bibinfo
  {pages} {261803} (\bibinfo {year} {2021})},\ \Eprint
  {http://arxiv.org/abs/2110.06096} {arXiv:2110.06096 [hep-ex]} \BibitemShut
  {NoStop}%
\bibitem [{\citenamefont {Csaki}\ \emph {et~al.}(2010)\citenamefont {Csaki},
  \citenamefont {Shirman},\ and\ \citenamefont {Terning}}]{Csaki:2010rv}%
  \BibitemOpen
  \bibfield  {author} {\bibinfo {author} {\bibfnamefont {C.}~\bibnamefont
  {Csaki}}, \bibinfo {author} {\bibfnamefont {Y.}~\bibnamefont {Shirman}}, \
  and\ \bibinfo {author} {\bibfnamefont {J.}~\bibnamefont {Terning}},\ }\href
  {\doibase 10.1103/PhysRevD.81.125028} {\bibfield  {journal} {\bibinfo
  {journal} {Phys. Rev. D}\ }\textbf {\bibinfo {volume} {81}},\ \bibinfo
  {pages} {125028} (\bibinfo {year} {2010})},\ \Eprint
  {http://arxiv.org/abs/1003.0448} {arXiv:1003.0448 [hep-th]} \BibitemShut
  {NoStop}%
\bibitem [{\citenamefont {Yang}\ \emph {et~al.}(2024)\citenamefont {Yang},
  \citenamefont {Gao},\ and\ \citenamefont {Peng}}]{Yang:2022uil}%
  \BibitemOpen
  \bibfield  {author} {\bibinfo {author} {\bibfnamefont {Q.}~\bibnamefont
  {Yang}}, \bibinfo {author} {\bibfnamefont {Y.}~\bibnamefont {Gao}}, \ and\
  \bibinfo {author} {\bibfnamefont {Z.}~\bibnamefont {Peng}},\ }\href {\doibase
  10.1038/s42005-024-01770-y} {\bibfield  {journal} {\bibinfo  {journal}
  {Commun. Phys.}\ }\textbf {\bibinfo {volume} {7}},\ \bibinfo {pages} {277}
  (\bibinfo {year} {2024})},\ \Eprint {http://arxiv.org/abs/2201.08291}
  {arXiv:2201.08291 [hep-ph]} \BibitemShut {NoStop}%
\bibitem [{\citenamefont {Yang}\ and\ \citenamefont {Di}(2018)}]{Yang:2016zaz}%
  \BibitemOpen
  \bibfield  {author} {\bibinfo {author} {\bibfnamefont {Q.}~\bibnamefont
  {Yang}}\ and\ \bibinfo {author} {\bibfnamefont {H.}~\bibnamefont {Di}},\
  }\href {\doibase 10.1016/j.physletb.2018.03.045} {\bibfield  {journal}
  {\bibinfo  {journal} {Phys. Lett. B}\ }\textbf {\bibinfo {volume} {780}},\
  \bibinfo {pages} {622} (\bibinfo {year} {2018})},\ \Eprint
  {http://arxiv.org/abs/1606.01492} {arXiv:1606.01492 [hep-ph]} \BibitemShut
  {NoStop}%
\bibitem [{\citenamefont {Dixit}\ \emph {et~al.}(2021)\citenamefont {Dixit},
  \citenamefont {Chakram}, \citenamefont {He}, \citenamefont {Agrawal},
  \citenamefont {Naik}, \citenamefont {Schuster},\ and\ \citenamefont
  {Chou}}]{Dixit:2020ymh}%
  \BibitemOpen
  \bibfield  {author} {\bibinfo {author} {\bibfnamefont {A.~V.}\ \bibnamefont
  {Dixit}}, \bibinfo {author} {\bibfnamefont {S.}~\bibnamefont {Chakram}},
  \bibinfo {author} {\bibfnamefont {K.}~\bibnamefont {He}}, \bibinfo {author}
  {\bibfnamefont {A.}~\bibnamefont {Agrawal}}, \bibinfo {author} {\bibfnamefont
  {R.~K.}\ \bibnamefont {Naik}}, \bibinfo {author} {\bibfnamefont {D.~I.}\
  \bibnamefont {Schuster}}, \ and\ \bibinfo {author} {\bibfnamefont
  {A.}~\bibnamefont {Chou}},\ }\href {\doibase 10.1103/PhysRevLett.126.141302}
  {\bibfield  {journal} {\bibinfo  {journal} {Phys. Rev. Lett.}\ }\textbf
  {\bibinfo {volume} {126}},\ \bibinfo {pages} {141302} (\bibinfo {year}
  {2021})},\ \Eprint {http://arxiv.org/abs/2008.12231} {arXiv:2008.12231
  [hep-ex]} \BibitemShut {NoStop}%
\bibitem [{\citenamefont {Braine}\ \emph {et~al.}(2020)\citenamefont {Braine}
  \emph {et~al.}}]{ADMX:2019uok}%
  \BibitemOpen
  \bibfield  {author} {\bibinfo {author} {\bibfnamefont {T.}~\bibnamefont
  {Braine}} \emph {et~al.} (\bibinfo {collaboration} {ADMX}),\ }\href {\doibase
  10.1103/PhysRevLett.124.101303} {\bibfield  {journal} {\bibinfo  {journal}
  {Phys. Rev. Lett.}\ }\textbf {\bibinfo {volume} {124}},\ \bibinfo {pages}
  {101303} (\bibinfo {year} {2020})},\ \Eprint
  {http://arxiv.org/abs/1910.08638} {arXiv:1910.08638 [hep-ex]} \BibitemShut
  {NoStop}%
\bibitem [{\citenamefont {O'Hare}(2020)}]{AxionLimits}%
  \BibitemOpen
  \bibfield  {author} {\bibinfo {author} {\bibfnamefont {C.}~\bibnamefont
  {O'Hare}},\ }\href {\doibase 10.5281/zenodo.3932430} {\enquote {\bibinfo
  {title} {cajohare/axionlimits: Axionlimits},}\ }\bibinfo {howpublished}
  {\url{https://cajohare.github.io/AxionLimits/}} (\bibinfo {year}
  {2020})\BibitemShut {NoStop}%
\bibitem [{\citenamefont {Duan}\ \emph {et~al.}(2023)\citenamefont {Duan},
  \citenamefont {Gao}, \citenamefont {Ji}, \citenamefont {Sun}, \citenamefont
  {Yao},\ and\ \citenamefont {Zhang}}]{Duan:2022nuy}%
  \BibitemOpen
  \bibfield  {author} {\bibinfo {author} {\bibfnamefont {J.}~\bibnamefont
  {Duan}}, \bibinfo {author} {\bibfnamefont {Y.}~\bibnamefont {Gao}}, \bibinfo
  {author} {\bibfnamefont {C.-Y.}\ \bibnamefont {Ji}}, \bibinfo {author}
  {\bibfnamefont {S.}~\bibnamefont {Sun}}, \bibinfo {author} {\bibfnamefont
  {Y.}~\bibnamefont {Yao}}, \ and\ \bibinfo {author} {\bibfnamefont {Y.-L.}\
  \bibnamefont {Zhang}},\ }\href {\doibase 10.1103/PhysRevD.107.015019}
  {\bibfield  {journal} {\bibinfo  {journal} {Phys. Rev. D}\ }\textbf {\bibinfo
  {volume} {107}},\ \bibinfo {pages} {015019} (\bibinfo {year} {2023})},\
  \Eprint {http://arxiv.org/abs/2206.13543} {arXiv:2206.13543 [hep-ph]}
  \BibitemShut {NoStop}%
\bibitem [{\citenamefont {Phipps}\ \emph {et~al.}(2020)\citenamefont {Phipps}
  \emph {et~al.}}]{Phipps:2019cqy}%
  \BibitemOpen
  \bibfield  {author} {\bibinfo {author} {\bibfnamefont {A.}~\bibnamefont
  {Phipps}} \emph {et~al.},\ }\href {\doibase 10.1007/978-3-030-43761-9_16}
  {\bibfield  {journal} {\bibinfo  {journal} {Springer Proc. Phys.}\ }\textbf
  {\bibinfo {volume} {245}},\ \bibinfo {pages} {139} (\bibinfo {year}
  {2020})},\ \Eprint {http://arxiv.org/abs/1906.08814} {arXiv:1906.08814
  [astro-ph.CO]} \BibitemShut {NoStop}%
\bibitem [{\citenamefont {Crisosto}\ \emph {et~al.}(2020)\citenamefont
  {Crisosto}, \citenamefont {Sikivie}, \citenamefont {Sullivan}, \citenamefont
  {Tanner}, \citenamefont {Yang},\ and\ \citenamefont
  {Rybka}}]{Crisosto:2019fcj}%
  \BibitemOpen
  \bibfield  {author} {\bibinfo {author} {\bibfnamefont {N.}~\bibnamefont
  {Crisosto}}, \bibinfo {author} {\bibfnamefont {P.}~\bibnamefont {Sikivie}},
  \bibinfo {author} {\bibfnamefont {N.~S.}\ \bibnamefont {Sullivan}}, \bibinfo
  {author} {\bibfnamefont {D.~B.}\ \bibnamefont {Tanner}}, \bibinfo {author}
  {\bibfnamefont {J.}~\bibnamefont {Yang}}, \ and\ \bibinfo {author}
  {\bibfnamefont {G.}~\bibnamefont {Rybka}},\ }\href {\doibase
  10.1103/PhysRevLett.124.241101} {\bibfield  {journal} {\bibinfo  {journal}
  {Phys. Rev. Lett.}\ }\textbf {\bibinfo {volume} {124}},\ \bibinfo {pages}
  {241101} (\bibinfo {year} {2020})},\ \Eprint
  {http://arxiv.org/abs/1911.05772} {arXiv:1911.05772 [astro-ph.CO]}
  \BibitemShut {NoStop}%
\bibitem [{\citenamefont {Godfrey}\ \emph {et~al.}(2021)\citenamefont {Godfrey}
  \emph {et~al.}}]{Godfrey:2021tvs}%
  \BibitemOpen
  \bibfield  {author} {\bibinfo {author} {\bibfnamefont {B.}~\bibnamefont
  {Godfrey}} \emph {et~al.},\ }\href {\doibase 10.1103/PhysRevD.104.012013}
  {\bibfield  {journal} {\bibinfo  {journal} {Phys. Rev. D}\ }\textbf {\bibinfo
  {volume} {104}},\ \bibinfo {pages} {012013} (\bibinfo {year} {2021})},\
  \Eprint {http://arxiv.org/abs/2101.02805} {arXiv:2101.02805
  [physics.ins-det]} \BibitemShut {NoStop}%
\bibitem [{\citenamefont {Levine}\ \emph {et~al.}(2024)\citenamefont {Levine},
  \citenamefont {Godfrey}, \citenamefont {Tyson}, \citenamefont {Tripathi},
  \citenamefont {Polin}, \citenamefont {Aminaei}, \citenamefont {Kolner},\ and\
  \citenamefont {Stucky}}]{Levine:2024noa}%
  \BibitemOpen
  \bibfield  {author} {\bibinfo {author} {\bibfnamefont {J.}~\bibnamefont
  {Levine}}, \bibinfo {author} {\bibfnamefont {B.}~\bibnamefont {Godfrey}},
  \bibinfo {author} {\bibfnamefont {J.~A.}\ \bibnamefont {Tyson}}, \bibinfo
  {author} {\bibfnamefont {S.~M.}\ \bibnamefont {Tripathi}}, \bibinfo {author}
  {\bibfnamefont {D.}~\bibnamefont {Polin}}, \bibinfo {author} {\bibfnamefont
  {A.}~\bibnamefont {Aminaei}}, \bibinfo {author} {\bibfnamefont {B.~H.}\
  \bibnamefont {Kolner}}, \ and\ \bibinfo {author} {\bibfnamefont
  {P.}~\bibnamefont {Stucky}},\ }\href@noop {} {\  (\bibinfo {year} {2024})},\
  \Eprint {http://arxiv.org/abs/2405.20444} {arXiv:2405.20444 [hep-ex]}
  \BibitemShut {NoStop}%
\bibitem [{\citenamefont {Preskill}\ \emph {et~al.}(1983)\citenamefont
  {Preskill}, \citenamefont {Wise},\ and\ \citenamefont
  {Wilczek}}]{Preskill:1982cy}%
  \BibitemOpen
  \bibfield  {author} {\bibinfo {author} {\bibfnamefont {J.}~\bibnamefont
  {Preskill}}, \bibinfo {author} {\bibfnamefont {M.~B.}\ \bibnamefont {Wise}},
  \ and\ \bibinfo {author} {\bibfnamefont {F.}~\bibnamefont {Wilczek}},\ }\href
  {\doibase 10.1016/0370-2693(83)90637-8} {\bibfield  {journal} {\bibinfo
  {journal} {Phys. Lett. B}\ }\textbf {\bibinfo {volume} {120}},\ \bibinfo
  {pages} {127} (\bibinfo {year} {1983})}\BibitemShut {NoStop}%
\bibitem [{\citenamefont {Abbott}\ and\ \citenamefont
  {Sikivie}(1983)}]{Abbott:1982af}%
  \BibitemOpen
  \bibfield  {author} {\bibinfo {author} {\bibfnamefont {L.~F.}\ \bibnamefont
  {Abbott}}\ and\ \bibinfo {author} {\bibfnamefont {P.}~\bibnamefont
  {Sikivie}},\ }\href {\doibase 10.1016/0370-2693(83)90638-X} {\bibfield
  {journal} {\bibinfo  {journal} {Phys. Lett. B}\ }\textbf {\bibinfo {volume}
  {120}},\ \bibinfo {pages} {133} (\bibinfo {year} {1983})}\BibitemShut
  {NoStop}%
\bibitem [{\citenamefont {Dine}\ and\ \citenamefont
  {Fischler}(1983)}]{Dine:1982ah}%
  \BibitemOpen
  \bibfield  {author} {\bibinfo {author} {\bibfnamefont {M.}~\bibnamefont
  {Dine}}\ and\ \bibinfo {author} {\bibfnamefont {W.}~\bibnamefont
  {Fischler}},\ }\href {\doibase 10.1016/0370-2693(83)90639-1} {\bibfield
  {journal} {\bibinfo  {journal} {Phys. Lett. B}\ }\textbf {\bibinfo {volume}
  {120}},\ \bibinfo {pages} {137} (\bibinfo {year} {1983})}\BibitemShut
  {NoStop}%
\bibitem [{\citenamefont {Long}\ and\ \citenamefont
  {Wang}(2019)}]{Long:2019lwl}%
  \BibitemOpen
  \bibfield  {author} {\bibinfo {author} {\bibfnamefont {A.~J.}\ \bibnamefont
  {Long}}\ and\ \bibinfo {author} {\bibfnamefont {L.-T.}\ \bibnamefont
  {Wang}},\ }\href {\doibase 10.1103/PhysRevD.99.063529} {\bibfield  {journal}
  {\bibinfo  {journal} {Phys. Rev. D}\ }\textbf {\bibinfo {volume} {99}},\
  \bibinfo {pages} {063529} (\bibinfo {year} {2019})},\ \Eprint
  {http://arxiv.org/abs/1901.03312} {arXiv:1901.03312 [hep-ph]} \BibitemShut
  {NoStop}%
\end{thebibliography}%

\end{document}